\documentclass[nofootinbib,preprintnumbers]{revtex4}

\usepackage[english]{babel}
\usepackage{graphicx}
\usepackage{psfrag}
\usepackage{amsmath}
\usepackage{amssymb}
\usepackage{verbatim}

\newcommand{\be}{\begin{equation}}
\newcommand{\ee}{\end{equation}}
\newcommand{\bea}{\begin{eqnarray}}
\newcommand{\eea}{\end{eqnarray}}
\newcommand{\bean}{\begin{eqnarray*}}
\newcommand{\eean}{\end{eqnarray*}}

\renewcommand{\b}{\langle}
\newcommand{\ket}{\rangle}

\newcommand{\irm}{{\rm i}}
\newcommand{\e}{{\rm e}}

\renewcommand{\d}{{\rm d}}
\newcommand{\cl}[1]{{\mathcal #1}}

\newcommand{\ts}{\textstyle}
\newcommand{\sst}{\scriptstyle}
\newcommand{\ssst}{\scriptscriptstyle}

\newcommand{\bZ}{\mathbb{Z}}

\newcommand{\clS}{\cl{S}}

\newcommand{\eq}[1]{(\ref{#1})}

\newcommand{\fig}[1]{Fig.\ \ref{#1}}

\newcommand{\tr}{{\rm tr}}

\newcommand{\pic}[4]
{
 \begin{figure}
 \begin{center}
 \includegraphics[height=#3]{#4}
 \end{center}
 \caption{\label{#1} #2}
 \end{figure}
}

\newcommand{\qed}{\nobreak \ifvmode \relax \else
      \ifdim\lastskip<1.5em \hskip-\lastskip
      \hskip1.5em plus0em minus0.5em \fi \nobreak
      \vrule height0.75em width0.5em depth0.25em\fi}

\newcommand{\sixj}[6]{
\left\{
\begin{array}{ccc}
#1 & #2 & #3 \\ 
#4 & #5 & #6
\end{array}
\right\}
}

\newcommand{\threejm}[6]{
\left(
\begin{array}{ccc}
#1 & #2 & #3 \\ 
#4 & #5 & #6
\end{array}
\right)
}

\newcommand{\Tt}{\tilde{T}}

\begin{document}

\title{Dual representation of Polyakov loop \\ in 3d SU(2) lattice Yang-Mills theory}
\author{Florian Conrady}
\email{fconrady@perimeterinstitute.ca}
\affiliation{Institute for Gravitational Physics and Geometry, Physics Department, Penn State University, University Park, Pennsylvania, U.S.A}
\preprint{IGPG-07/6-7}

\begin{abstract}
We consider the expectation value of a Polyakov loop in 3d SU(2) lattice Yang--Mills theory and transform it to the dual representation in terms of sums over spins. The spin dependence of the amplitudes is computed explicitly by a graphical method. We also determine the asymptotic (large spin) limit of the amplitude factors.
\end{abstract}

\maketitle

\section{Introduction}
\label{introduction}

Strong--coupling expansions of lattice gauge theory provide an appealing physical model for the large--distance part of the quark potential 
\cite{Wilsonconfinement}: the sum over strong--coupling graphs corresponds to fluctuating strings of flux and naturally explains the confining potential 
and its $N$--ality dependence. The problem is that these sums are hard to analyze in the continuum limit: they are either not convergent at weak coupling, or they are convergent, but difficult to evaluate. As the coupling decreases, a growing number of increasingly complex graphs has to be summed.

Here, we will deal with the latter type of strong--coupling expansion, i.e.\ the one which converges for strong and weak coupling. It results from an expansion of plaquette actions into a basis of characters, and from a subsequent integration over the connection \cite{Munsterhightemperature,DrouffeZuber}. Thus, the sum over graphs is not an expansion in powers of $\beta$, but rather a \textit{dual representation} that is equivalent to the original lattice gauge theory \cite{Anishettyetal,HallidaySuranyi,DiakonovPetrov,OecklPfeifferdualofpurenonAbelian}. It can be viewed as a resummation of an expansion in $\beta$. 
For this reason, we try to avoid the adjective ``strong--coupling'' and call the graphs instead \textit{spin foams} \cite{OecklPfeifferdualofpurenonAbelian}. Originally, this name was introduced for SU(2) \cite{Baezspinfoammodels}, but it is also used for general gauge groups. In the case of SU(2), one obtains a sum over spin assignments to the lattice that satisfy certain spin coupling conditions. Each admissible configuration is a spin foam.

In the early literature on the strong--coupling expansion, weighting factors were only determined for spin foams up to a certain complexity. 
Later, amplitudes were analyzed in general, and given as a function of arbitrary spin foams. For the partition function of 3d SU(2) lattice Yang--Mills theory, the form of the complete sum was first determined by Anishetty, Cheluvaraja, Sharatchandra and Mathur \cite{Anishettyetal}. For the same theory, Diakonov and Petrov computed in some detail the dual transform of a Wilson loop \cite{DiakonovPetrov}. In an abstract form, Oeckl and Pfeiffer wrote down the dual representation for any dimension $d\ge 2$, any compact gauge group $G$ and for general observables \cite{OecklPfeifferdualofpurenonAbelian}. 

In this paper, we present two new results in this direction: we use a graphical scheme to determine, in complete detail, the dual transform for a Polaykov loop in 3d SU(2) Yang--Mills theory. The graphical method makes the derivation more transparent and easier to check than a purely algebraic calculation. This result is used in ref.\ \cite{ConradyKhavkinestringrepresentation} to derive an exact string representation for two Polyakov loops.

The second result concerns the asymptotic limit of the amplitudes for large spins. The large spin limit is important for understanding weakly coupled processes in the dual representation. As was suggested by Diakonov \& Petrov \cite{DiakonovPetrov}, and further explored by the author \cite{ConradyglumonII}, dual gluons arise as spin waves in a weak--coupling approximation. This was shown by an argument that employs the large spin limit of $6j$--symbols. 

In the presence of a Wilson loop, the argument is incomplete, however, since the amplitudes involve also $9j$--symbols, and for these the asymptotic behaviour is not known so far \cite{ConradyglumonII}. We improve this situation with the second result of the paper: we determine the large spin limit of the amplitudes for the Polyakov loop. This is possible, since we have chosen a zig--zag path for the loop, for which the amplitudes factorize into $6j$--symbols. 

The paper is organized as follows: in section \ref{SU2latticegaugetheory}, we specify 3d SU(2) lattice Yang--Mills theory with the heat kernel action. 
In section \ref{dualrepresentationofpartitionfunction} and \ref{dualrepresentationofPolyakovloop}, we describe the dual transform of the partition function and of the expectation value of a Polyakov. The derivation is given in the appendix.

\section{SU(2) lattice Yang-Mills theory in 3 dimensions}
\label{SU2latticegaugetheory}

The partition function of 3--dimensional SU(2) lattice Yang-Mills theory is defined by a path integral over SU(2)--valued link (or edge) variables $U_e$ on a cubic lattice $\kappa$ with periodic boundary conditions:
\be
\label{partitionfunction}
Z = \int\left({\ts\prod\limits_{e\subset\kappa}}\d U_e\right) \exp\Big(-\sum_f \clS_f(U_f)\Big)
\ee
The face (or plaquette) action $\clS_f$ depends on the holonomy $U_f$ around the face. We choose $S_f$ to be the heat kernel action (for more details on the definition, see \cite{MenottiOnofri}). The heat kernel action has a particularly simple expansion in terms of characters, namely, 
\be
\exp\Big(- \clS_f(U_f)\Big) = \sum_j\;(2j+1)\,\e^{-\frac{2}{\beta}\,j(j + 1)}\,\chi_j(U_f)\,.
\ee
The coupling factor $\beta$ is related to the gauge coupling $g$ via
\be
\beta = \frac{4}{a g^2} + \frac{1}{3}\,.
\ee
The expectation value of a Polyakov loop $C$ in the representation $j$ is
\be
\label{Polykovloopexpectationvalue}
\b \tr_j U_C\ket = \int\left({\ts\prod\limits_{e\subset\kappa}}\d U_e\right)\; \tr_j U_C\,\exp\Big(-\sum_f \clS_f(U_f)\Big)\,.
\ee
$U_C$ denotes the holonomy along the loop $C$.

\section{Dual representation of partition function}
\label{dualrepresentationofpartitionfunction}

\begin{figure}
\setlength{\unitlength}{1cm}
\begin{center}
(a)\quad \parbox{6.6cm}{\includegraphics[height=5cm]{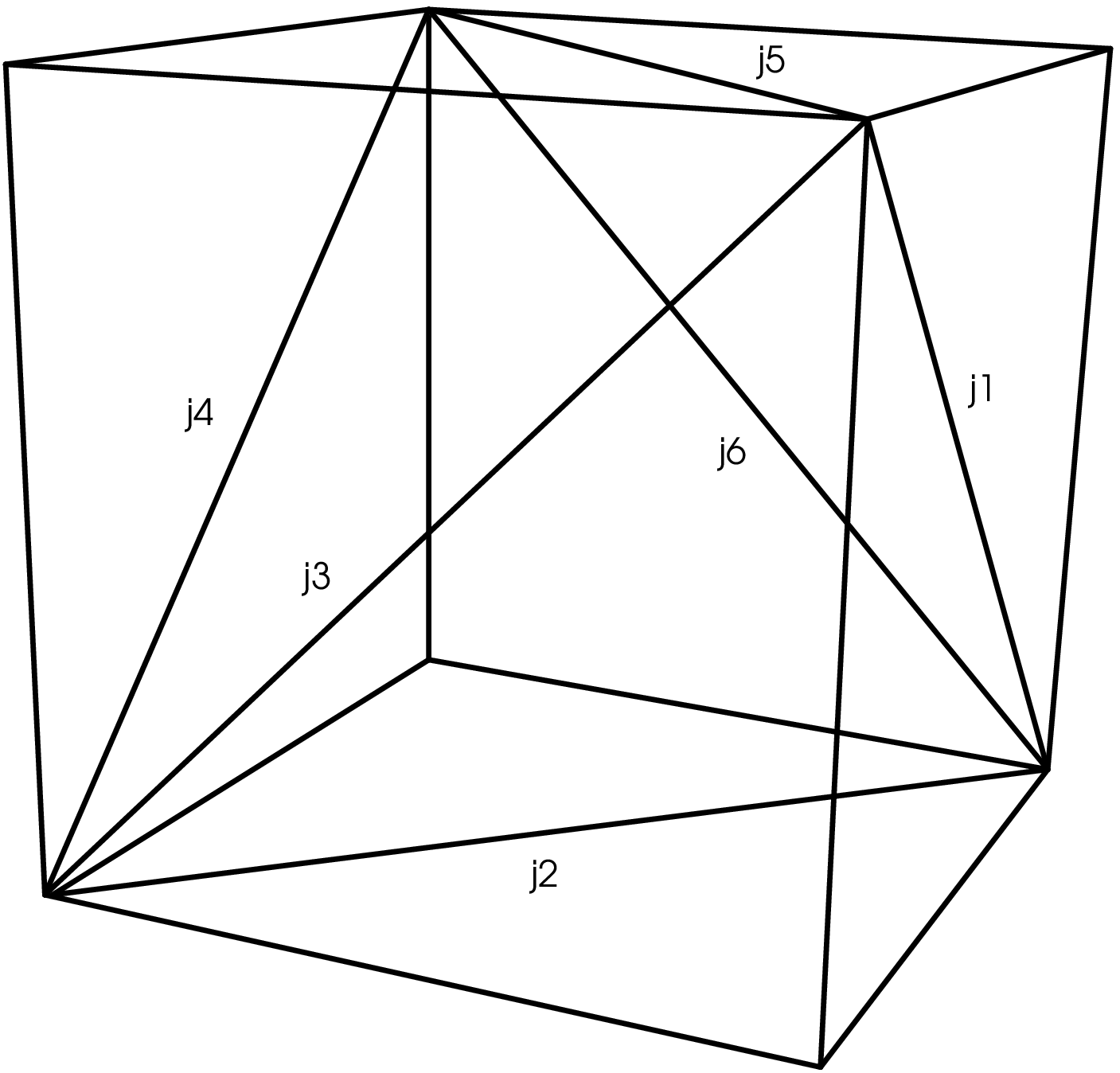}}\hspace{1cm}
(b)\quad \parbox{6.6cm}{\includegraphics[height=5cm]{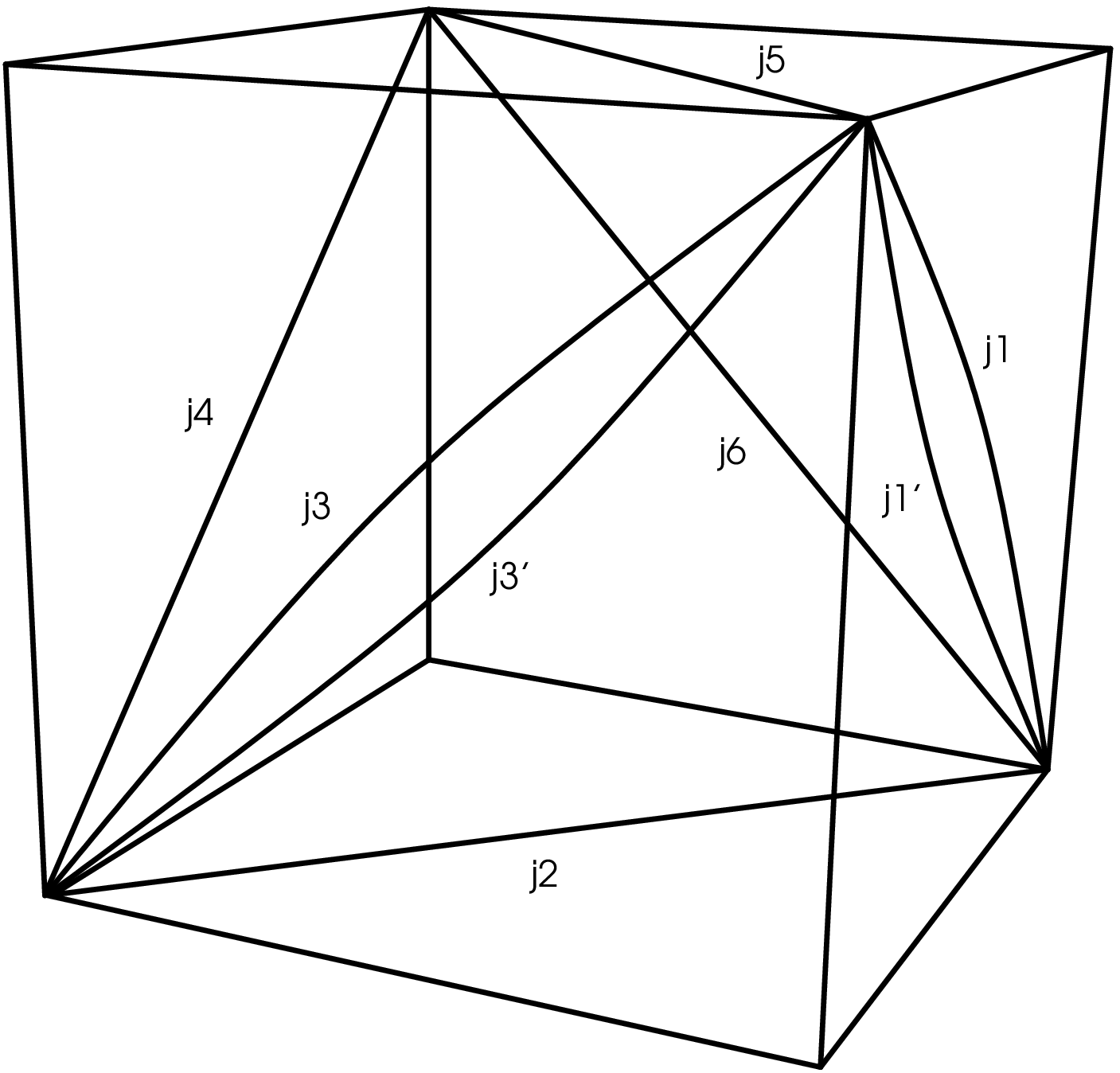}}
\end{center}
\caption{\label{TandTprime} 
(a) Triangulation $T$ that is obtained by adding diagonals to cubes of $\kappa^*$. 
(b) Modified triangulation $\Tt$ with double edges where the Polyakov line passes through.}
\end{figure}

Let us split the set of cubes into two subsets, called white and black, or even and odd, so that we obtain a 3-dimensional ``checkerboard''. Then, go to the dual lattice $\kappa^*$, and call vertices even if they are dual to even cubes, and odd if they are dual to odd cubes. Connect all odd vertices by edges. The resulting new complex is a triangulation $T$ (see \fig{TandTprime}a). 

\psfrag{j1}{$j_1$}
\psfrag{j2}{$j_2$}
\psfrag{j3}{$j_3$}
\psfrag{j4}{$j_4$}
\psfrag{j5}{$j_5$}
\psfrag{j6}{$j_6$}
\psfrag{j3'}{$j'_3$}
\psfrag{j1'}{$j'_1$}

With these conventions, the dual transform of the partition function takes the following form:
\be
\label{spinfoamsumpartitionfunctionT}
Z = \sum_{\{j_e\}_T}
\left(\prod_{e\subset T} (2j_e+1)\right)
\left(\prod_{t\subset T} A_t\right)
\left(\prod_{e\subset\kappa^*}\;(-1)^{2j_e}\,\e^{-\frac{2}{\beta}\,j_e(j_e + 1)}\right)
\ee
Each configuration $\{j_e\}_T$ is an assignment of spins $j_e$ to edges $e$ of $T$ such that for each triangle of $T$ the spins satisfy the triangle inequality. The edges of $T$ belong to two groups: edges that are identified with edges in $\kappa^*$, and diagonal edges that were added to $\kappa^*$ in order to form the triangulation $T$. In the amplitude, every edge contributes with the dimension $2j_e + 1$ of the representaiton $j_e$. In addition, edges of $\kappa^*$ give a sign factor and an exponential of the Casimir of the representation. For each tetrahedron, we get an amplitude factor
\be
\label{amplitudetetrahedron}
A_t\quad =\quad \sixj{j_1}{j_2}{j_3}{j_4}{j_5}{j_6}\,,
\ee
where the spins $j_1$, $j_2$ and $j_3$ are read off from any triangle in the tetrahedron. The spins $j_4$, $j_5$, and $j_6$ are the spins on the edges opposing those of $j_1$, $j_2$ and $j_3$. 

For large spins, $6j$--symbols are approximated by the Ponzano--Regge formula \cite{PonzanoRegge}
\be
A_t \approx \frac{1}{\sqrt{6\pi V_t}}\left(\e^{\irm R_t} + \e^{-\irm R_t}\right)\,.
\label{asymptotics}
\ee
Here, $V_t$ is the volume of the tetrahedron $t$ when its edges have length $j_e + 1/2$, and $R_t$ is its contribution to the Regge action:
\be
R_t = \sum_{e \subset t} \left(j_e + 1/2\right)\theta_{te} + \frac{\pi}{4}
\ee 
If we assume that large spins dominate, we can use this to obtain an asymptotic version of the spin foam sum:
\be
Z = \sum_{\{j_e\}_T}\sum_{\{s_t\}}
\left(\prod_{e\subset T} (2j_e+1)\right)
\left(\prod_{t\subset T} \frac{1}{\sqrt{6\pi V_t}} \exp\Big(\irm s_t R_t\Big)\right)
\exp\left(-\irm\sum_{e\subset\kappa^*} 2\pi j_e 
- \sum_{e\subset\kappa^*}\frac{2}{\beta}\,j_e(j_e + 1)\right)
\ee
We sum over signs $s_t = \pm 1$ for each tetrahedron, due to the two terms of opposite phase in formula \eq{asymptotics}.

\section{Dual representation of Polyakov loop}
\label{dualrepresentationofPolyakovloop}

\psfrag{Polyakov loop}{Polyakov loop}
\psfrag{C}{$C$}
\begin{figure}
\setlength{\unitlength}{1cm}
\begin{center}
(a)\quad\parbox{7.8cm}{\includegraphics[height=4.5cm]{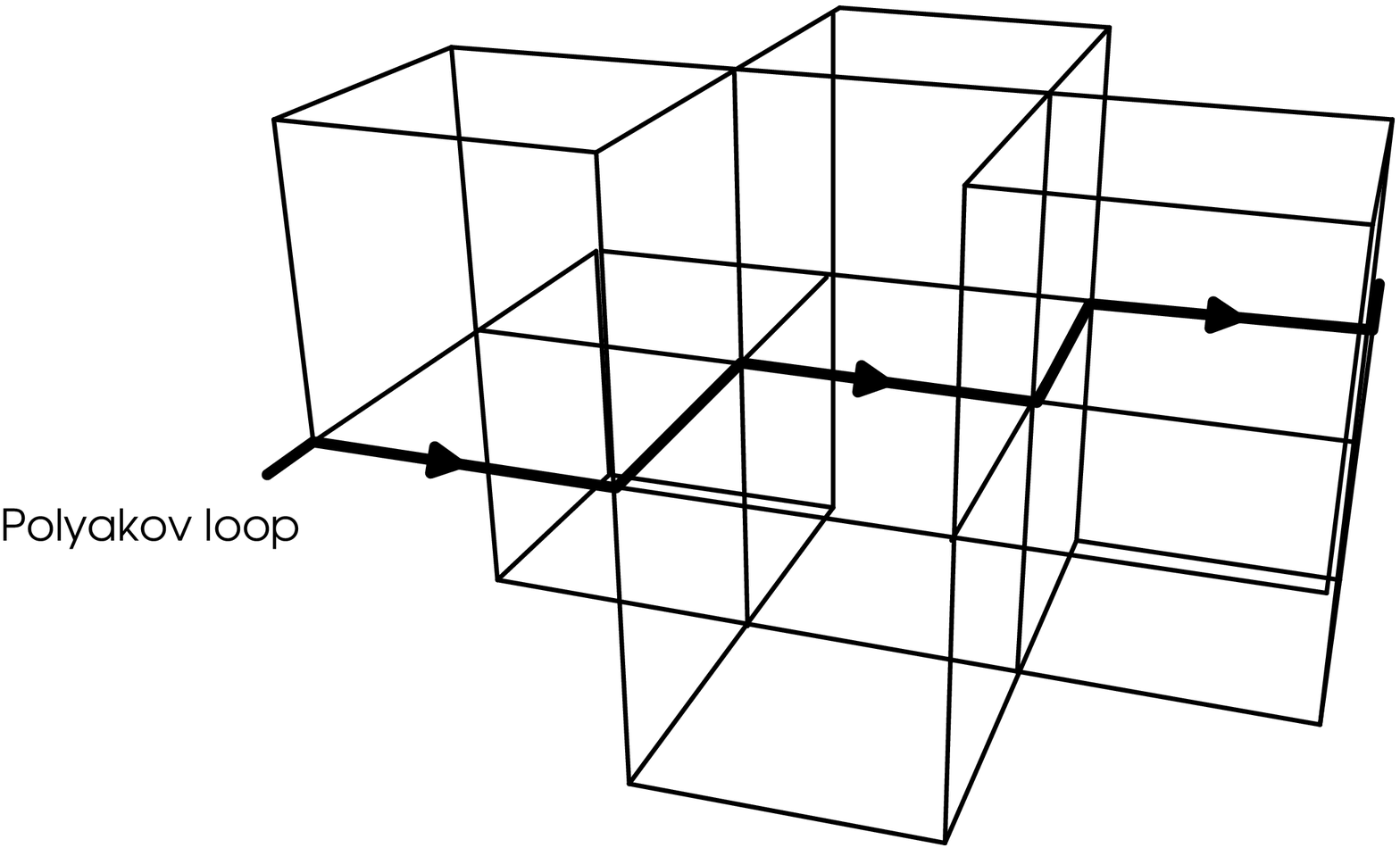}}
\hspace{1cm}
(b)\quad\parbox{5.6cm}{\includegraphics[height=5.5cm]{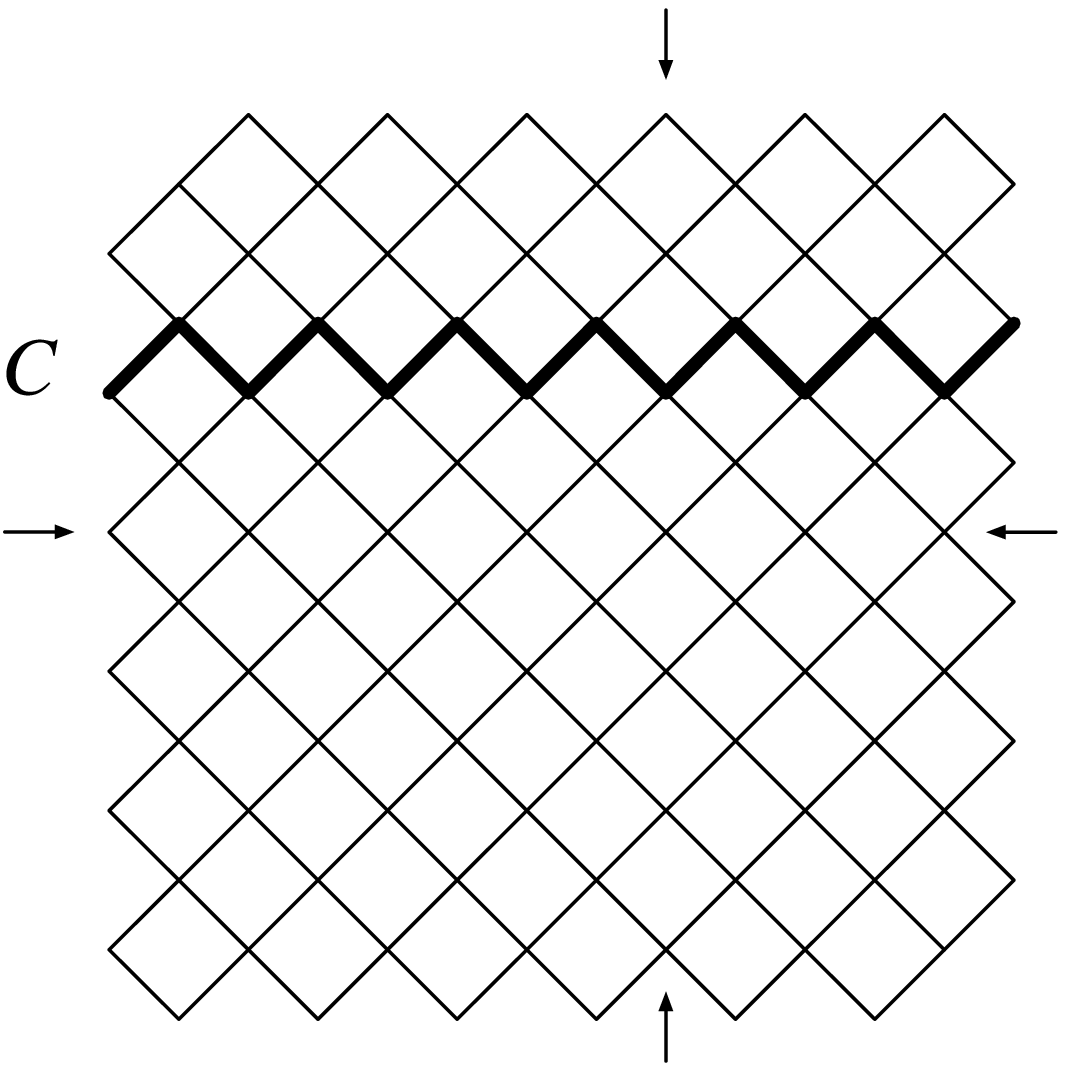}}
\end{center}
\caption{\label{Polyakovloop} 
(a) Zig--zag path of a Polyakov loop through the lattice $\kappa$. Only even cubes are shown. (b) The Polyakov loop in a 2d slice of the lattice $\kappa$. The arrows indicate how lattice points are identified.}
\end{figure}

\renewcommand{\arraystretch}{1.5}
We now come to the dual transform of a Polyakov loop. We choose a Polyakov loop that runs along a zig--zig path through the lattice, and adopt suitable boundary conditions  (see \fig{Polyakovloop}a and b). More precisely, we assume that the Polyakov loop runs through a
1--2--plane of the lattice, and that the sites of the lattice are given by points
\be
(x_1,x_2,x_3) \quad=\quad 
\left\{
\begin{array}{c}
\left(\sqrt{2}(k,l),m\right) \\
\mbox{or} \\
\left(\sqrt{2}(k-\frac{1}{2},l-\frac{1}{2}),m\right)
\end{array}
\right. \qquad\mbox{for $(k,l,m)\in \bZ^3$}
\ee
with boundary conditions
\be
(x_1 + \sqrt{2}L, x_2,x_3) \quad\simeq\quad (x_1,x_2 + \sqrt{2}L,x_3) \quad\simeq\quad (x_1,x_2,x_3+L) \quad\simeq\quad (x_1,x_2,x_3)\,.
\ee
That is, in each 1--2--plane, we identify points on opposing ends of diagonals, and in the $x_3$--direction we use ordinary periodic boundary conditions.
\renewcommand{\arraystretch}{1}

The zig--zag path is not essential for determining the dual transform, but it helps us to find the asymptotic limit of the amplitudes. If we used straight lines, the amplitudes would contain $9j$--amplitudes, and for these the asymptotic form is not known so far. For zig--zag paths, on the other hand, the amplitudes reduce to $6j$--symbols and we can use the formulas by Ponzano \& Regge and Edmonds. 

To describe the sum over spin foams, we have to modify the triangulation $T$: consider all faces of the dual lattice $\kappa^*$ which are dual to edges of the Polyakov loop. In each of these faces, we add a second diagonal edge. The resulting complex is the modified triangulation $\Tt$ (see \fig{TandTprime}b).

Then, the dual representation can be specified as follows:
\be
\label{spinfoamsumPolyakovloop}
\b \tr_j U_C\ket = \frac{1}{Z}\,\sum_{\{j_e\}_{\Tt}}
\left(\prod_{e\subset \Tt} (2j_e+1)\right)
\left(\prod_{t\subset \Tt} A_t\right)
\left(\prod_{e\subset\kappa^*}\;(-1)^{2j_e}\,\e^{-\frac{2}{\beta}\,j_e(j_e + 1)}\right)\,.
\ee
Each configuration $\{j_e\}_{\Tt}$ is an assignment of spins $e$ to edges of $\Tt$ such that 1.\ for each triangle of $\Tt$ the spins satisfy the triangle inequality, and 2.\ at every double edge the spins satisfy the inequality
\be
\left|j_i - j\right| \le j'_i \le j_i + j\,.
\label{triangleinequalitywithj}
\ee
Compared to the amplitude in \eq{spinfoamsumpartitionfunctionT}, there is only one difference: in every cube of the dual lattice $\kappa^*$ where the Polyakov loop passes through, we get an additional amplitude factor $A_t$. To understand this, consider \fig{TandTprime}b: here, the Polyakov loop enters through the front of the cube, and exits on the right side, so we have added two edges---an edge with spin $j'_3$ on the front and an edge with spin $j'_1$ on the right side. Due to the presence of these two edges, we can define an additional \textit{degenerate} tetrahedron in the cube. Apart from the tetrahedron defined by the edges $j_1$, $j_2$, $j_3$, $j_4$, $j_5$ and $j_6$, we get a tetrahedron formed by the edges $j_1$, $j_3$, $j_2$, $j'_3$ and $j'_1$, where the sixth edge is shrunk to zero length.

\begin{figure}
\setlength{\unitlength}{1cm}
\begin{center}
(a)\quad \parbox{5cm}{\includegraphics[height=4cm]{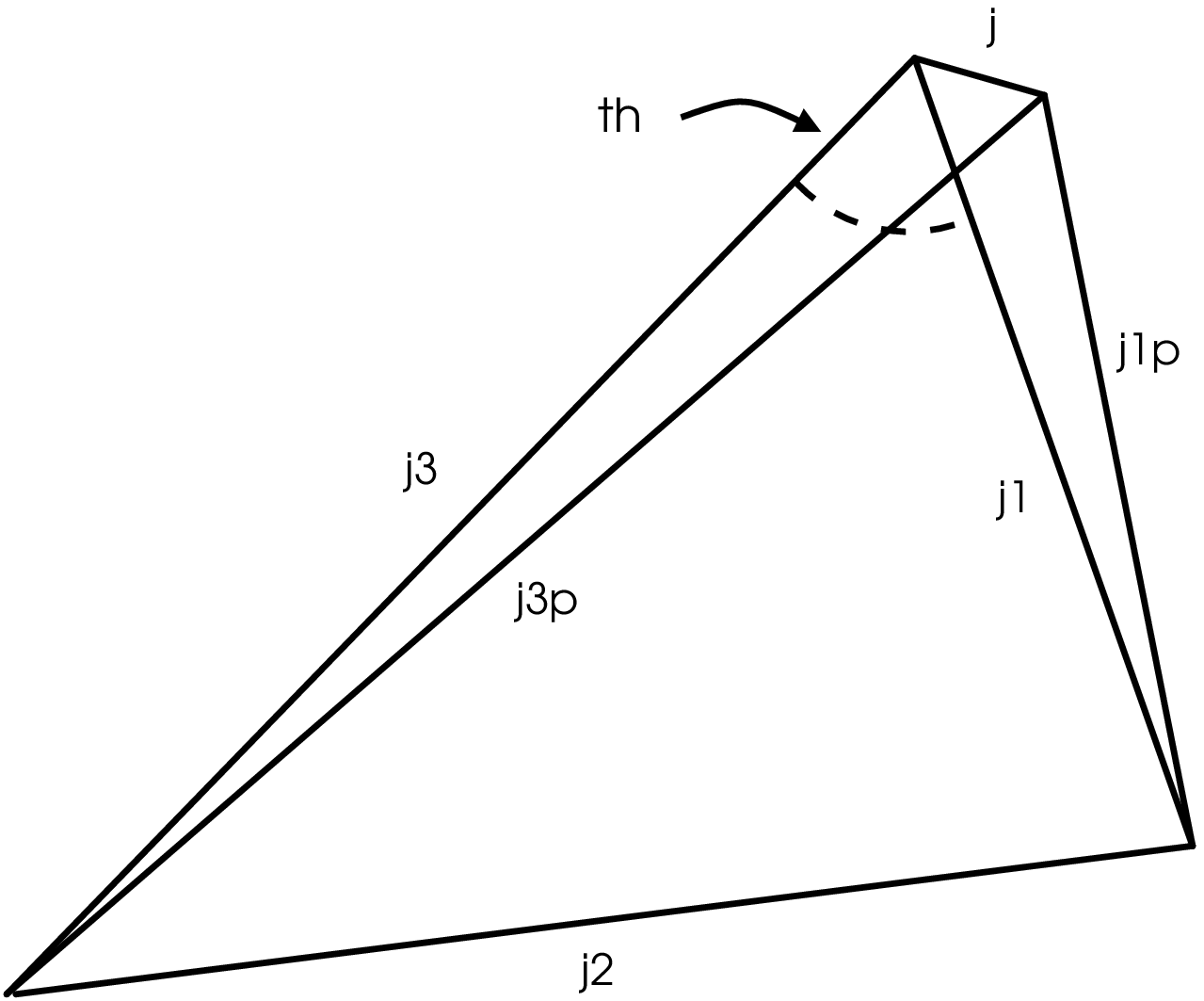}}\hspace{1cm}
(b)\quad \parbox{9cm}{\includegraphics[height=5.5cm]{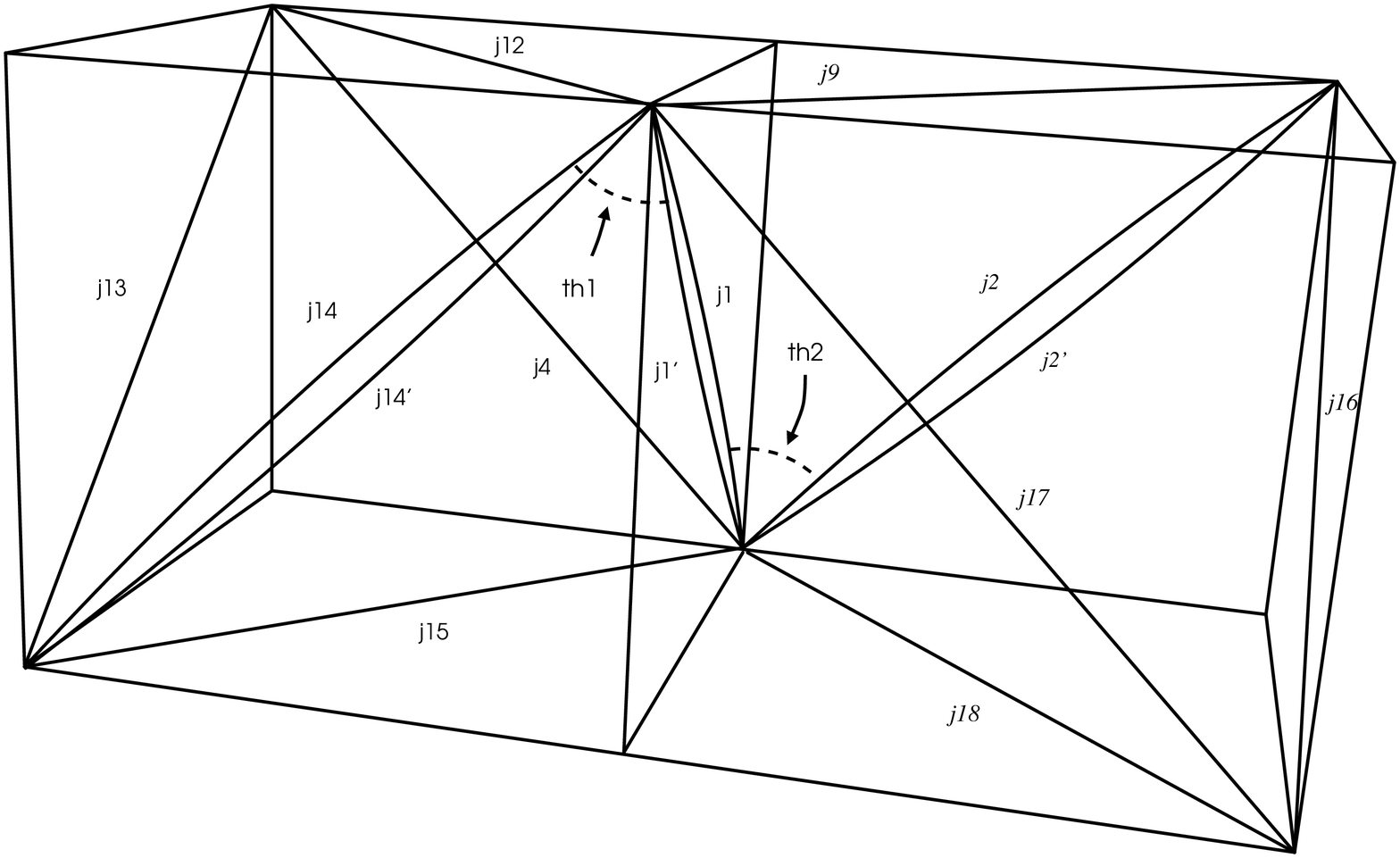}}
\end{center}
\caption{\label{extendedcubesvertexab} 
(a) Extended tetrahedron that results from adding an edge of spin $j$ to a degenerate tetrahedron of $\Tt$. 
(b) Two consecutive degenerate tetrahedra $t_1$ and $t_2$ in the modififed triangulation $\Tt$.}
\end{figure}

Thus, we have two tetrahedra in the middle of the cube: the one we already had in \fig{TandTprime}a and another, degenerate one. For this second tetrahedron, we receive an additional amplitude factor $A_t$ which is defined as follows: take the degenerate tetrahdron, and extend it by inserting an edge of spin $j$ at the corner where the two double edges meet (see \fig{extendedcubesvertexab}a). To this tetrahedron, we associate the factor 
\be
A_t = (-1)^{j_3 - j'_3}\,(-1)^{j_1 - j'_1}\,(-1)^{j_1 + j_2 + j_3 + j}\,\sixj{j_1}{j_3}{j_2}{j'_3}{j'_1}{j}\,.
\label{amplitudedegeneratetetrahedron}
\ee
The spins $j_1$, $j_2$ and $j_3$ are read off from one of the two triangles not containing $j$: if the short edge with spin $j$ is drawn at the top (as in \fig{extendedcubesvertexab}a), this triangle is on the left side of $j$ in the direction of passage of the Polyakov loop, i.e.\ on the left side in the direction from $j_3$, $j'_3$ towards $j_1$, $j'_1$. 

\psfrag{j1}{$j_1$}
\psfrag{j2}{$j_2$}
\psfrag{j3}{$j_3$}
\psfrag{j4}{$j_4$}
\psfrag{j5}{$j_5$}
\psfrag{j6}{$j_6$}
\psfrag{j3p}{$j'_3$}
\psfrag{j1p}{$j'_1$}
\psfrag{j}{$j$}
\psfrag{th}{$\vartheta_t$}

\psfrag{j1}{$j_1$}
\psfrag{j1'}{$j_1'$}
\psfrag{j2}{$j_2$}
\psfrag{j2'}{$j_2'$}
\psfrag{j3}{$j_3$}
\psfrag{j4}{$j_4$}
\psfrag{j5}{$j_5$}
\psfrag{j6}{$j_6$}
\psfrag{j7}{$j_7$}
\psfrag{j8}{$j_8$}
\psfrag{j9}{$j_9$}
\psfrag{j10}{$j_{10}$}
\psfrag{j11}{$j_{11}$}
\psfrag{j12}{$j_{12}$}
\psfrag{j13}{$j_{13}$}
\psfrag{j14}{$j_{14}$}
\psfrag{j14'}{$j_{14}'$}
\psfrag{j15}{$j_{15}$}
\psfrag{j16}{$j_{16}$}
\psfrag{j17}{$j_{17}$}
\psfrag{j18}{$j_{18}$}
\psfrag{th1}{$\vartheta_{t_1}$}
\psfrag{th2}{$\vartheta_{t_2}$}

\setlength{\jot}{0.3cm}
With this additional rule, the spin foam sum \eq{spinfoamsumPolyakovloop} is completely specified. As in the case of the partition function, we can derive an asymptotic representation, assuming that all spins are large except the one of the Polyakov loop.
To deal with the asymptotics of the degenerate tetrahedra, we use Edmonds' formula \cite{Edmonds}:
\be
\sixj{j_1}{j_3}{j_2}{j'_3}{j'_1}{j} = \frac{(-1)^{j_1 + j_3 + j_2 + j + m}}{\sqrt{(2j_1+1)(2j_3+1)}}\, d^j_{mn}(\vartheta_t)\,, 
\ee
$d^j$ denotes the Wigner $d$-function in the representation $j$ and the angle $\vartheta_t$ is defined as the angle between the two edges connected to $j$ in the extended tetrahedron (see \fig{extendedcubesvertexab}a):
\be
\cos\vartheta_t = \frac{j_1(j_1+1) + j_3(j_3+1) - j_2(j_2+1)}{2\sqrt{j_1(j_1+1) j_3(j_3+1)}}
\ee
Let us apply this formula to two consecutive degenerate tetrahedra in $\Tt$ (see \fig{extendedcubesvertexab}b). They give the amplitudes
\bea
&& 
(-1)^{j'_1 + j'_2 + j_9 + j}\,\sixj{j'_1}{j'_2}{j_9}{j_2}{j_1}{j}\;(-1)^{j_1 + j_{14} + j_{15} + j}\,\sixj{j_1}{j_{14}}{j_{15}}{j'_{14}}{j'_1}{j} \\
&=& 
(-1)^{k+l}\,(-1)^{j_1 + j_2 + j_9 + j}\,\sixj{j_1}{j_2}{j_9}{j_2+k}{j_1+l}{j}\;(-1)^{j_1 + j_{14} + j_{15} + j}\,\sixj{j_1}{j_{14}}{j_{15}}{j_{14}+m}{j_1+l}{j} 
\nonumber \\
&=& 
(-1)^{k+l}\,\frac{(-1)^k}{\sqrt{(2j_2+1)(2j_1+1)}}\,d^j_{kl}(\vartheta_{t_2})\;\frac{(-1)^m}{\sqrt{(2j_{14}+1)(2j_1+1)}}\,d^j_{ml}(\vartheta_{t_1}) \\
&=& \frac{1}{\sqrt{2j_2+1} (2j_1+1) \sqrt{2j_{14}+1}}\,(-1)^{2k + m + l}\,d^j_{kl}(\vartheta_{t_2})\,(-1)^{m-l}\,d^j_{lm}(\vartheta_{t_1}) \\
&=& \frac{1}{\sqrt{2j_2+1} (2j_1+1) \sqrt{2j_{14}+1}}\,d^j_{kl}(\vartheta_{t_2})\,d^j_{lm}(\vartheta_{t_1})\,.
\eea
If we multiply these factors for the complete sequence of degenerate tetrahedra $t_i$ along the Polyakov line, the total result becomes
\bea
\lefteqn{\b \tr_j U_C\ket \approx
\frac{1}{Z}\,\sum_{\{j_e\}_{\Tt}}\sum_{\{s_t\}}
\left(\prod_{e\subset T} (2j_e+1)\right)\;
\left(\prod_i\;(-1)^{j_i - j'_i}\,d^j_{j'_{i+1} - j_{i+1},\, j'_i - j_i}(\vartheta_i)\right)} \\
&& \times\,
\left(\prod_{\mathrm{non-degenerate}\; t\subset \Tt} \frac{1}{\sqrt{6\pi V_t}} \exp\Big(\irm s_t R_t\Big)\right)
\exp\left(- \irm\sum_{e\subset\kappa^*} 2\pi j_e 
- \sum_{e\subset\kappa^*} \frac{2}{\beta}\,j_e(j_e + 1)\right)\,.
\label{asymptoticlimitPolyaykovloop}
\eea
In this expression, the sequence of double edges is numbered by $i = 1,\ldots, N$. The associated spins are denoted by $j_i$ and $j'_i$, and $\vartheta_i$ is the angle between edges number $i$ and $i+1$.

\section{Discussion}
\label{discussion}

In this paper, we computed the explicit dual transform of the expectation value of a Polyakov loop in 3d SU(2) lattice Yang--Mills theory.
To improve the transparency of the derivation, most calculations were done in a graphical scheme (see appendix). In ref.\ \cite{ConradyKhavkinestringrepresentation}, our result is used to derive an exact string representation of two Polyakov loops. Due to the specific choice of the loop---a zig--zag path---the amplitudes factorized into $6j$--symbols and we could determine their asymptotic large spin limit. Such a limit is useful when one tries to understand how spin waves and Coulomb force emerge in the dual representation (see ref.\ \cite{ConradyglumonII}). Our asymptotic expression may be also helpful for analyzing the sign behaviour of the amplitude, which is an important issue in Monte Carlo simulations \cite{ChristensenCherringtonKhavkine}.

As a side remark, we point out a similarity between the asymptotic formula \eq{asymptoticlimitPolyaykovloop} and the amplitude for spinning particles in 3d quantum gravity (see eq.\ (153) in appendix B.2 of \cite{FreidelLivinePonzanoReggerevisitedIII}). In formula \eq{asymptoticlimitPolyaykovloop}, the Polyakov loop appears as a product of Wigner $d$--symbols in the representation $j$. A similar structure is obtained when we consider a loop--shaped Feyman diagram of a spinning particle in 3d quantum gravity. Such a relation is not entirely surprising, since 3d SU(2) Yang--Mills theory can be regarded as a deformation of 3d quantum gravity (see e.g.\ \cite{ConradyglumonI,ConradyglumonII}).

\begin{acknowledgments}
We thank Wade Cherrington, Dan Christensen and Igor Khavkine for discussions. This work was supported in part by the NSF grant PHY-0456913 and the Eberly research funds.
\end{acknowledgments} 

\begin{appendix}

\setlength{\jot}{0.5cm}

\section{Graphical derivation of dual representation}
\label{graphicalderivationofdualrepresentation}

In this appendix we derive the dual representation of the expectation value \eq{Polykovloopexpectationvalue} by a graphical method.
First we will explain our graphical notation and express a number of identities in terms of it. Then, we apply these identities
to obtain the dual representation.

\subsection{Conventions}

There exist various graphical schemes for representating calculations with SU(2) tensors.
A standard refererence is the work by El--Baz \& Castel \cite{ElBazCastel}: it incorporates the earlier notation by Yutsis, Levinson \& Vanagas 
and Brink \& Satchler \cite{YutsisLevinsonVanagas,BrinkSatchler}, and extends it by symbols for representation matrices and group integrals. 
Our present choice of notation will be similar to these references as far as invariant tensors are concerned. 
For representation matrices and integrals, we prefer to use a different convention which is inspired by ref.\ \cite{GirelliOecklPerez}. 

A directed line with label $j$ represents the identity in the spin $j$ representation of SU(2):
\psfrag{m}{$m$}
\psfrag{m'}{$m'$}
\psfrag{j}{$j$}
\be
\parbox{5cm}{\includegraphics[height=0.8cm]{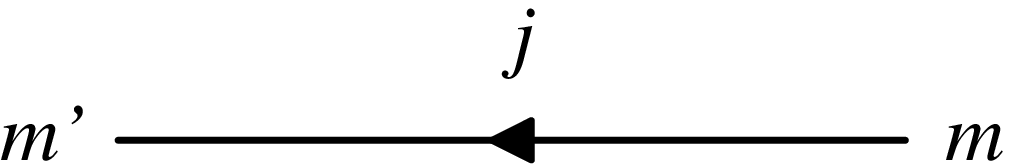}} \quad := \quad \delta^{m'}{}_m
\ee
Similarly, a representation matrix is symbolized by 
\be
\parbox{5cm}{\includegraphics[height=1.3cm]{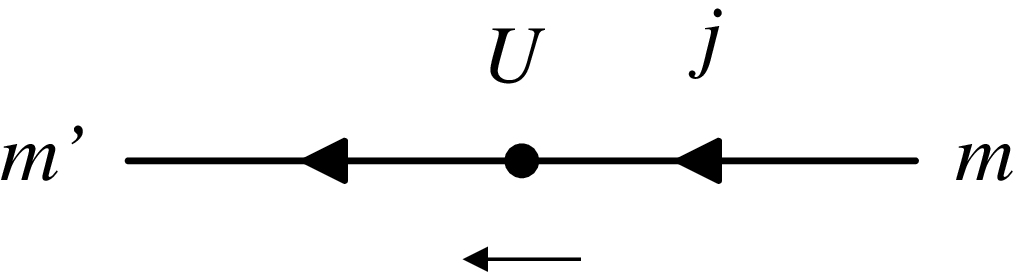}} \quad := \quad D^j(U)^{m'}{}_m
\ee
The target of the line corresponds to a vector in the $j$--representation $V_j$, while its source corresponds to a one--form in the dual $V^*_j$.
The dot indicates the dependence on the group element $U$, and the arrow on it specifies whether we have $D^j(U)$ or $D^j(U^{-1})$.
Integrations are indicated by ``cables'' around one or several lines:
\psfrag{m1}{$m_1$}
\psfrag{m'1}{$m'_1$}
\psfrag{m2}{$m_2$}
\psfrag{m'2}{$m'_2$}
\psfrag{m3}{$m_3$}
\psfrag{m'3}{$m'_3$}
\be
\parbox{5.1cm}{\includegraphics[height=1.9cm]{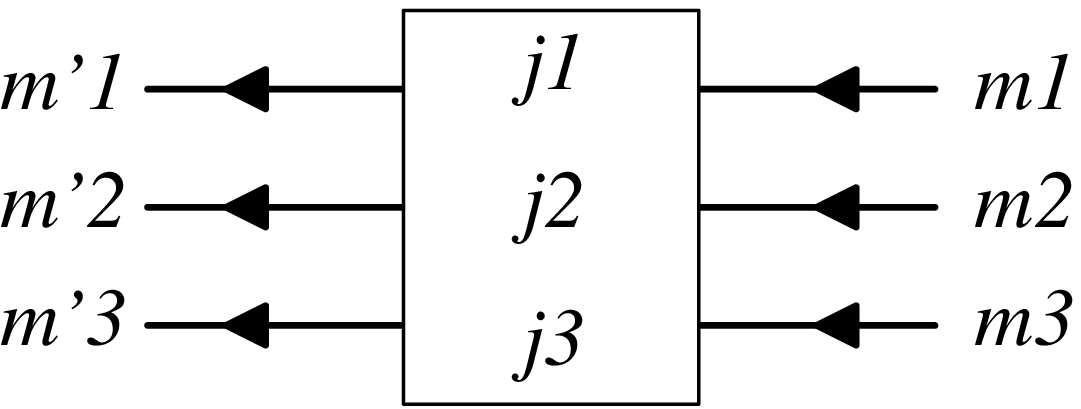}} \quad := \quad \int \d U\; D^{j_1}(U)^{m'_1}{}_{m_1}\,D^{j_2}(U)^{m'_2}{}_{m_2}\,D^{j_3}(U)^{m'_3}{}_{m_3}
\ee
We use $3jm$--symbols to construct normalized invariant tensors for triple tensor products of irreps. For such tensor products, the subspace of invariant tensors is one--dimensional, so we only need to fix one normalized tensor for each type of space. For the tensor product $V_{j_1}\otimes V_{j_2}\otimes V_{j_2}$ we take the $3jm$--symbol itself, and symbolize it by a node with three outgoing arrows:
\be
\parbox{3.3cm}{\includegraphics[height=2.8cm]{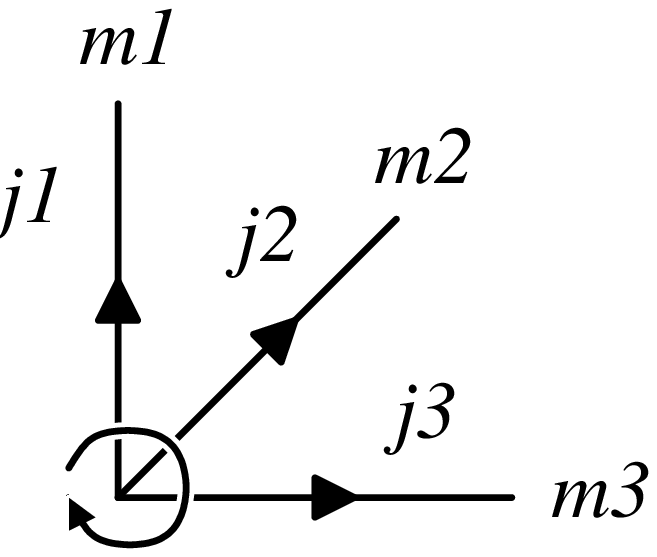}} \quad := \quad \threejm{j_1}{j_2}{j_3}{m_1}{m_2}{m_3}
\label{alloutward}
\ee
The order of spins is specified by a circle on the node. This is necessary, since permutations change the sign:
\be
\threejm{j_1}{j_2}{j_3}{m_1}{m_2}{m_3}\quad =\quad (-1)^{j_1 + j_2 + j_3}\threejm{j_1}{j_3}{j_2}{m_1}{m_3}{m_2}\,.
\ee 
For the dual space $V^*_{j_1}\otimes V^*_{j_2}\otimes V^*_{j_2}$, we choose the dual of \eq{alloutward}:
\be
\parbox{3.3cm}{\includegraphics[height=2.8cm]{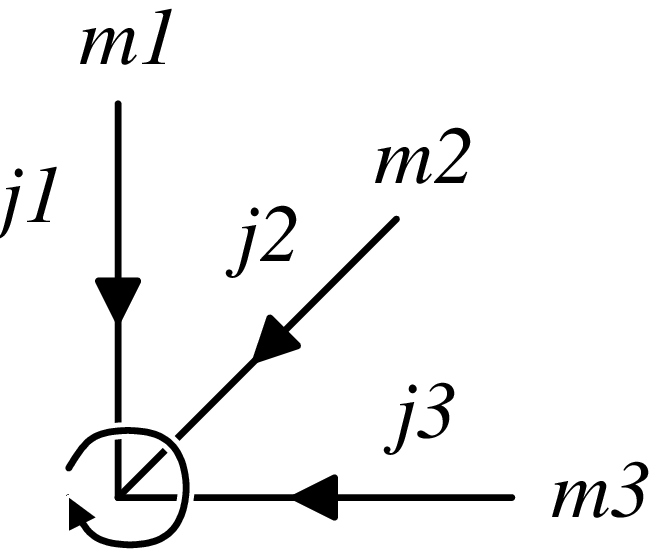}} \quad := \quad \left(\parbox{3.3cm}{\includegraphics[height=2.8cm]{10_threejm}}\right)^* 
\quad = \quad \threejm{j_1}{j_2}{j_3}{m_1}{m_2}{m_3}
\ee
The components are again those of the $3jm$--symbol, since the $3jm$--symbol is real. To construct normalized invariant tensors with upper and lower indices, we define the tensor
\psfrag{j'}{$j'$}
\psfrag{m'}{$m'$}
\psfrag{0}{$0$}
\bea
&& \parbox{5.5cm}{\includegraphics[height=1cm]{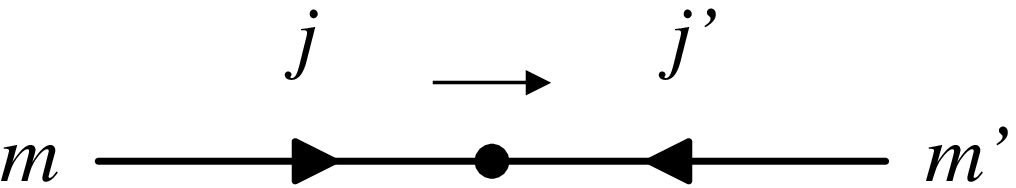}} \quad:=\quad 
\sqrt{2j_1 + 1}\times \parbox{4.9cm}{\includegraphics[height=2.3cm]{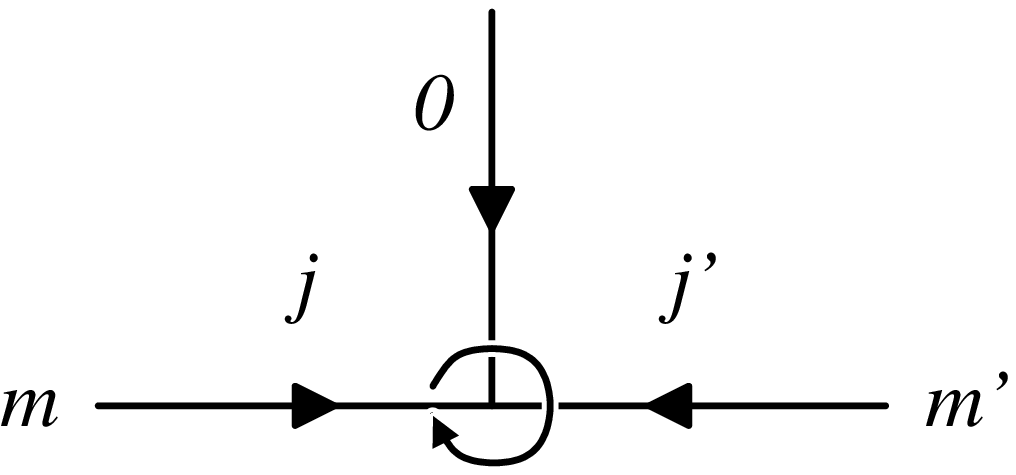}} \nonumber \\
&&= \sqrt{2j_1 + 1}\threejm{j}{0}{j'}{m}{0}{m'} = (-1)^{j+m}\,\delta_{m,-m'}\,\delta_{j,j'}
\eea
It is a normalized and invariant tensor in $V^*_j\otimes V^*_j$. By contracting it with the tensor \eq{alloutward} we obtain a normalized invariant tensor in $V_{j_1}\otimes V_{j_2}\otimes V^*_{j_3}$:
\bea
&& \parbox{3.3cm}{\includegraphics[height=2.8cm]{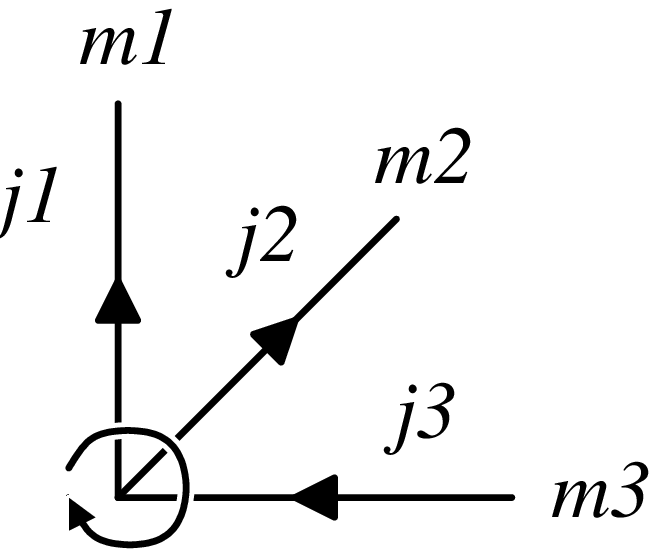}} \quad:=\quad 
\parbox{5.3cm}{\includegraphics[height=2.8cm]{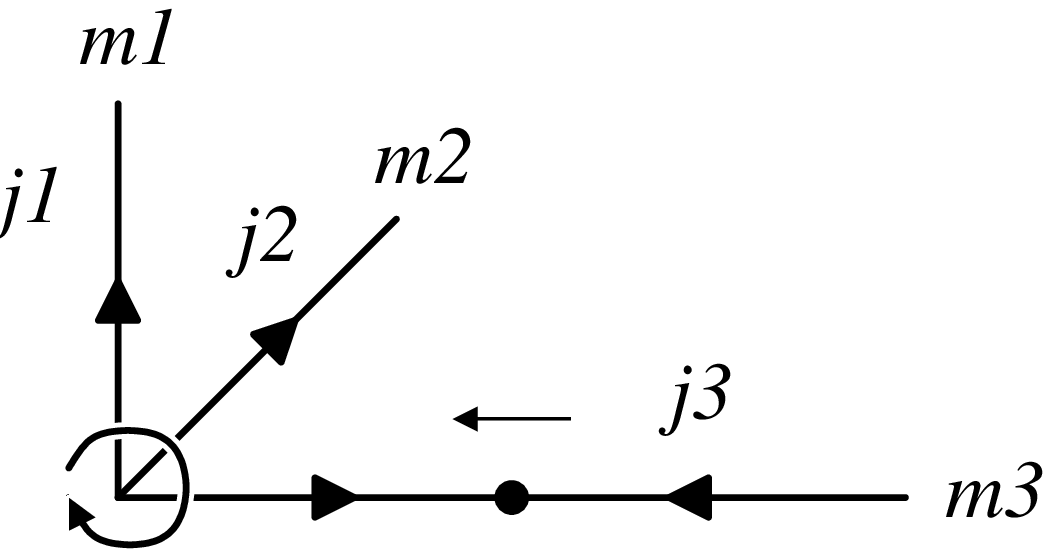}} \nonumber \\
&&= \sum_{m'_3} (-1)^{j_3 + m_3}\,\delta_{m_3, -m'_3} \threejm{j_1}{j_2}{j_3}{m_1}{m_2}{m'_3}  = (-1)^{j_3 + m_3} \threejm{j_1}{j_2}{j_3}{m_1}{m_2}{-m_3} 
\eea
The invariant tensor in $V^*_{j_1}\otimes V^*_{j_2}\otimes V_{j_3}$ results from dualization.
\be
\parbox{3.3cm}{\includegraphics[height=2.8cm]{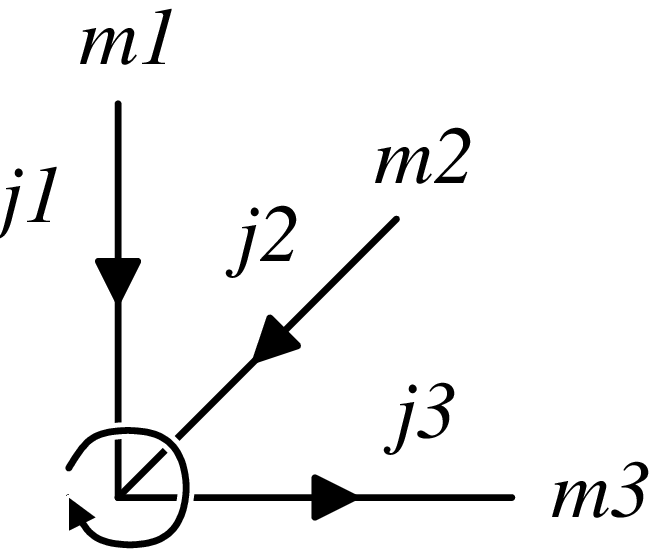}} \quad := \quad \left(\parbox{3.3cm}{\includegraphics[height=2.8cm]{14_threejmoneinward}}\right)^* 
\quad = \quad (-1)^{j_3 + m_3} \threejm{j_1}{j_2}{j_3}{m_1}{m_2}{-m_3} 
\ee
The remaining possibilities are fixed by applying the same procedure to $j_1$ or $j_2$ instead of $j_3$.

\subsection{Identities}

Next we express a number of identities in our graphical notation: we have that
\psfrag{ia}{$i_a$}
\psfrag{ja}{$j_a$}
\psfrag{ib}{$i_b$}
\psfrag{jb}{$j_b$}
\psfrag{JA}{$J_A$}
\psfrag{JB}{$J_B$}
\be
\parbox{5.4cm}{\includegraphics[height=2cm]{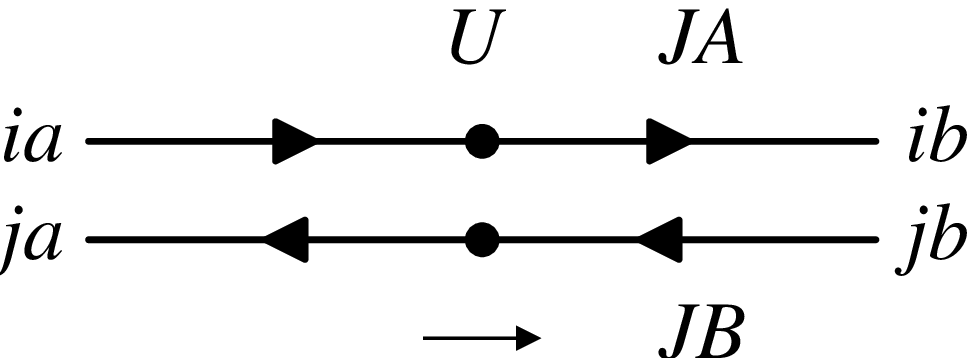}} \quad = \quad \sum_{j_1}\; (2j_1+1)\!\!\!\!\parbox{5.4cm}{\includegraphics[height=2.5cm]{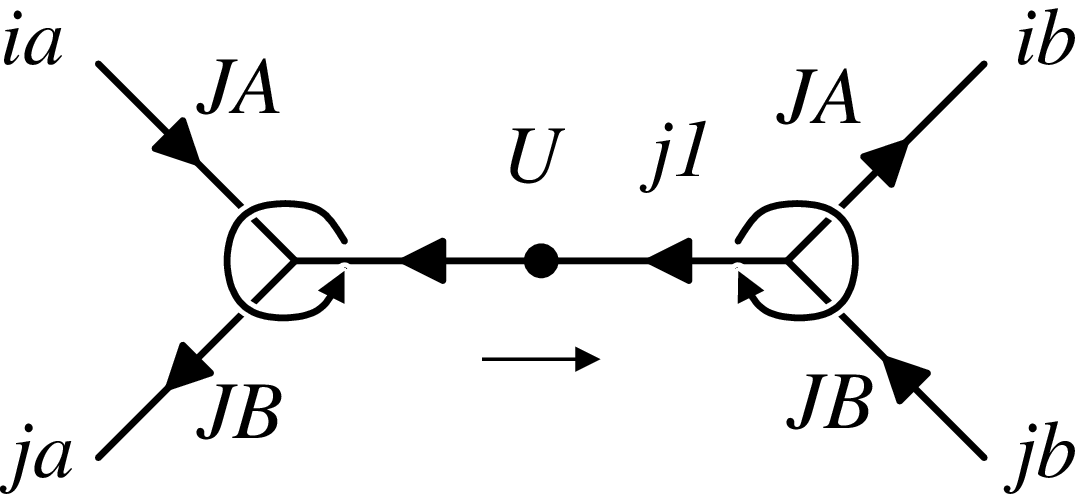}}\,,
\label{twostrands}
\ee
\bea
\lefteqn{D^{J_A}(U_1)^{i_b}{}_{i_a}\,D^{J_B}(U_1^{-1})^{j_a}{}_{j_b}} \nonumber \\ 
&&= \sum_{j_1}\; (2j_1+1)\,
(-1)^{J_B + j_b}\threejm{j_1}{J_A}{J_B}{o_b}{i_b}{-j_b}\,
(-1)^{J_B + j_a}\threejm{j_1}{J_A}{J_B}{o_a}{i_a}{-j_a} D^{j_1}(U_1^{-1})^{o_a}{}_{o_b}\,.
\eea
\psfrag{oa}{$o_a$}
\psfrag{ua}{$u_a$}
\psfrag{ob}{$o_b$}
\psfrag{ub}{$u_b$}
\psfrag{j1}{$j_1$}
\psfrag{j1'}{$j'_1$}
Integration over the group gives
\bea
\parbox{6cm}{\includegraphics[height=1.9cm]{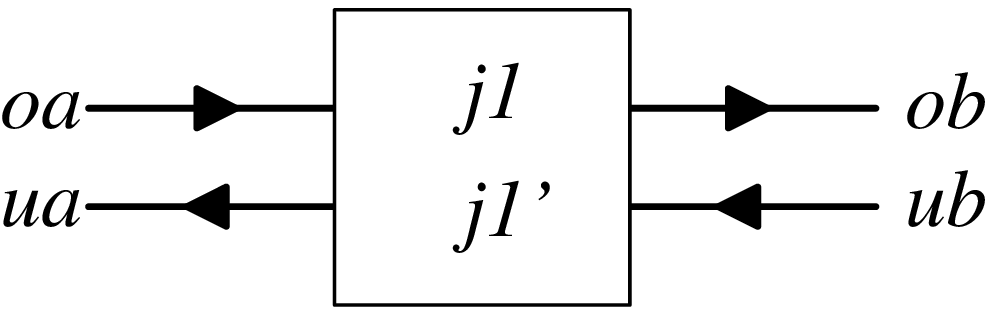}}
&=& 
\parbox{5.8cm}{\includegraphics[height=2.5cm]{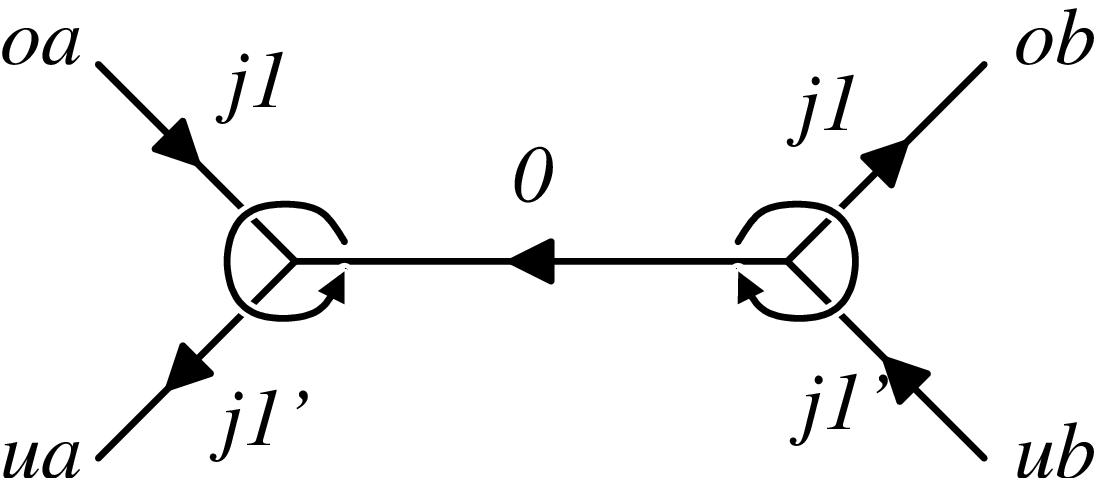}} \nonumber \\
&=& 
(2j_1+1)^{-1}\!\!\parbox{5.8cm}{\includegraphics[height=2.5cm]{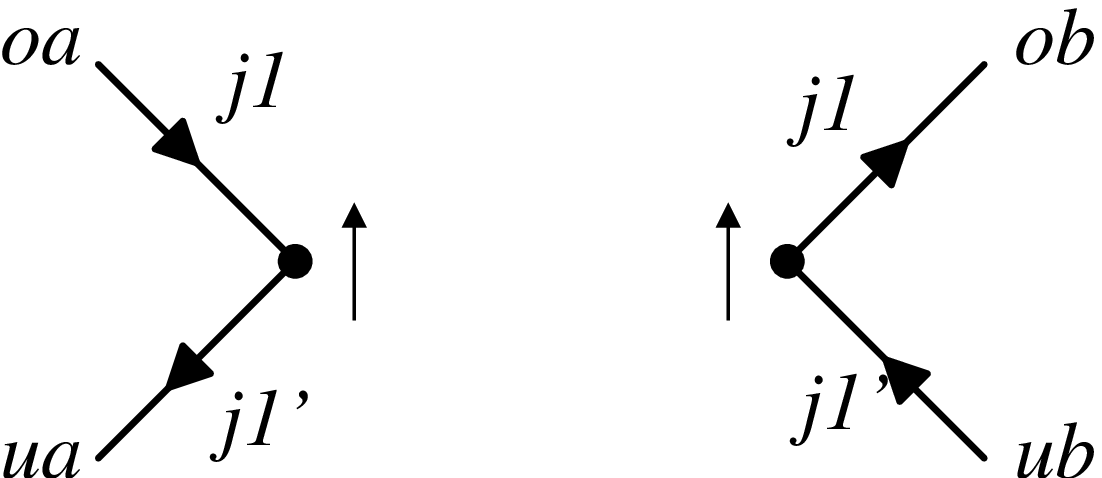}}\,,
\label{integral2strands}
\eea
\bea
\int \d U\; D^{j_1}(U)^{o_a}{}_{o_b}\,D^{j'_1}(U_1^{-1})^{u_a}{}_{u_b} &=& \threejm{j_1}{0}{j'_1}{u_a}{0}{o_a} \threejm{j_1}{0}{j'_1}{u_a}{0}{o_a} \nonumber \\
&=&
\frac{\delta_{j_1, j'_1}}{2j_1 + 1}\;\, (-1)^{j_1 + u_a}\,\delta_{u_a, -o_a}\; (-1)^{j_1 + u_b}\,\delta_{u_b, -o_b}\,.
\eea

\psfrag{m1'}{$m'_1$}
\psfrag{m2'}{$m'_2$}
\psfrag{m3'}{$m'_3$}
For three representation matrices, the integral identity reads
\be
\parbox{5.8cm}{\includegraphics[height=2.2cm]{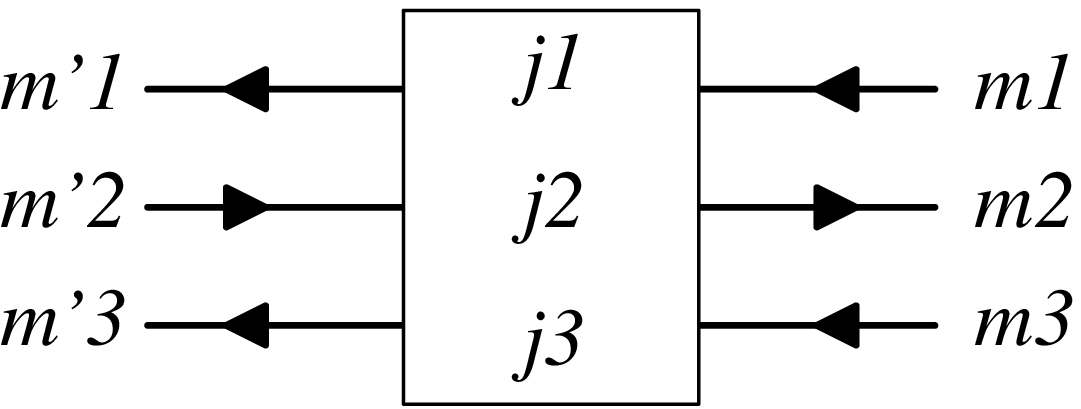}} \quad = \quad 
\parbox{6.1cm}{\includegraphics[height=2.5cm]{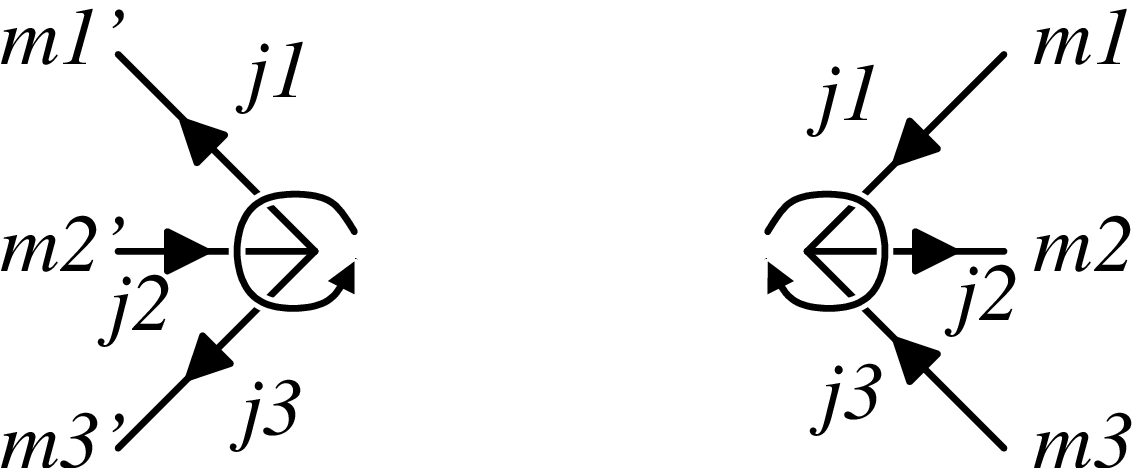}}\,,
\label{integral3strands}
\ee
\bea
\lefteqn{\int \d U\; D^{j_1}(U)^{m'_1}{}_{m_1}\,D^{j_2}(U^{-1})^{m_2}{}_{m'_2}\,D^{j_3}(U)^{m'_3}{}_{m_3}} \nonumber \\
&&=
(-1)^{j_2 + m'_2} \threejm{j_1}{j_2}{j_3}{m'_1}{-m'_2}{m'_3} (-1)^{j_2 + m_2} \threejm{j_1}{j_2}{j_3}{m_1}{-m_2}{m_3}\,.
\eea
The $6j$-symbol can be defined as a contraction of four $3jm$--symbols:
\bea
\lefteqn{\sixj{j_1}{j_2}{j_3}{j_4}{j_5}{j_6} \quad = \quad \parbox{5.3cm}{\includegraphics[height=3.2cm]{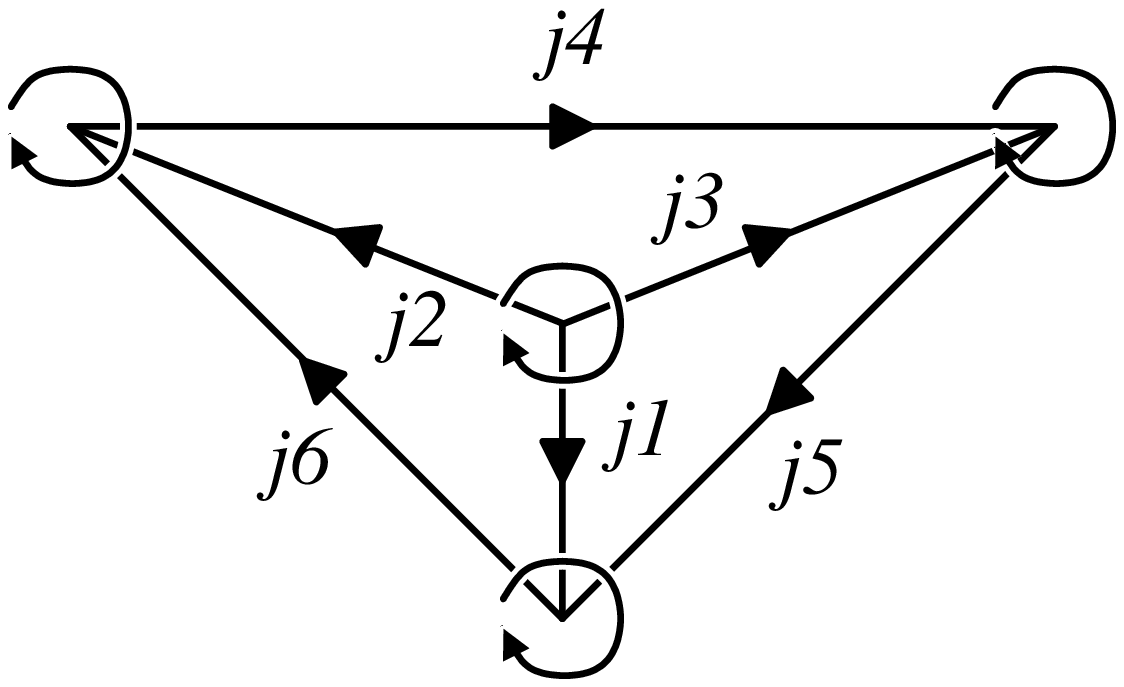}}} \\
&&= \sum_{klmnop} (-1)^{j_4+n+j_5+o+j_6+p}
\threejm{j_1}{j_2}{j_3}{k}{l}{m} \threejm{j_1}{j_5}{j_6}{k}{o}{-p} \threejm{j_2}{j_6}{j_4}{l}{p}{-n} \threejm{j_3}{j_4}{j_5}{m}{n}{-o}
\eea
This is equivalent to eq.\ (13), p.296, in Varshalovich et al.\ \cite{Varshalovich}. We also need the identity
\psfrag{j1}{$j_1$}
\psfrag{j2}{$j_2$}
\psfrag{j3}{$j_3$}
\psfrag{j4}{$j_4$}
\psfrag{j5}{$j_5$}
\psfrag{j6}{$j_6$}
\be
\parbox{4.3cm}{\includegraphics[height=4cm]{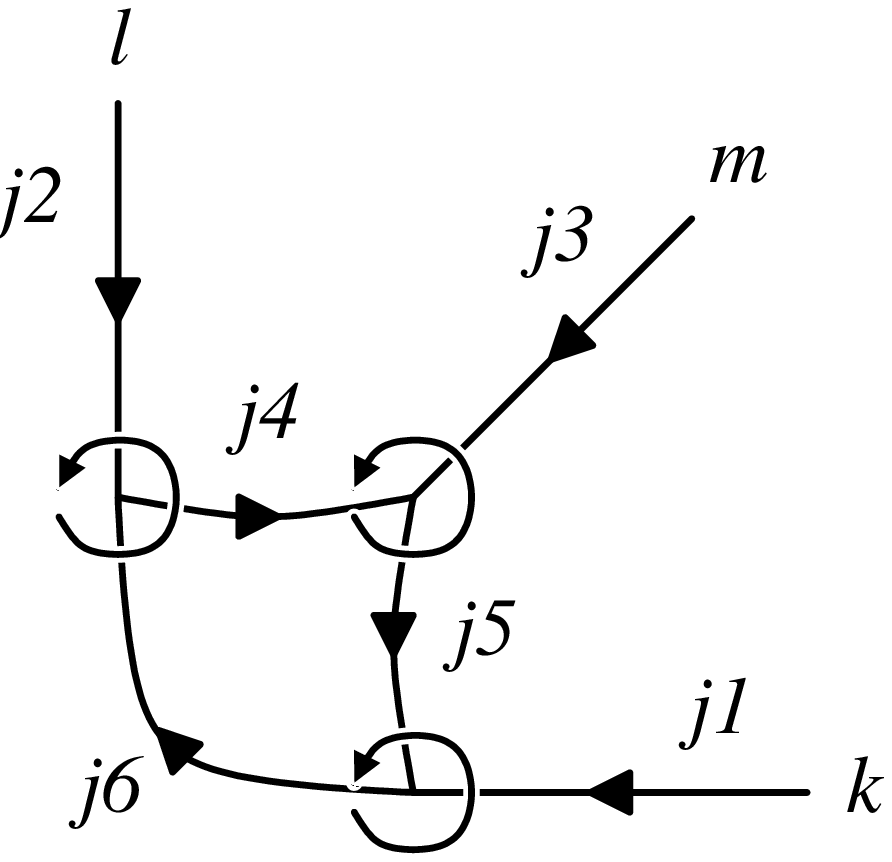}} \quad = \quad \parbox{5.3cm}{\includegraphics[height=3.2cm]{24_6j}}
\parbox{3.3cm}{\includegraphics[height=2.8cm]{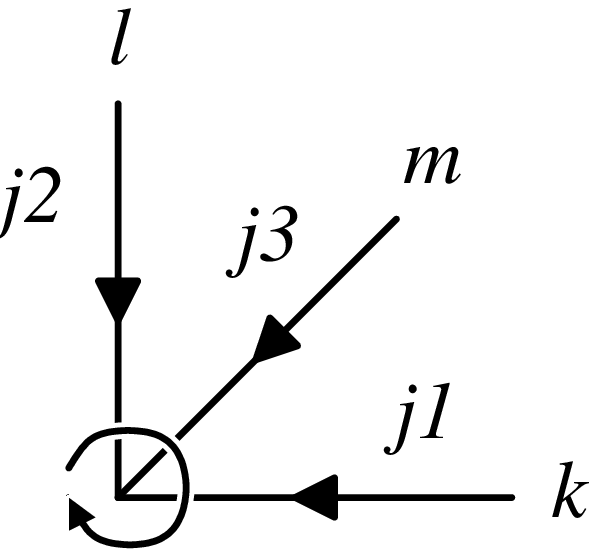}}\,,
\label{extractionidentity}
\ee
\be
\sum_{nop}\; (-1)^{j_6 + p + j_4 + n + j_5 + o} \threejm{j_1}{j_5}{j_6}{k}{o}{-p} \threejm{j_2}{j_6}{j_4}{l}{p}{-n} \threejm{j_3}{j_4}{j_5}{m}{n}{-o} 
= \sixj{j_1}{j_2}{j_3}{j_4}{j_5}{j_6} \threejm{j_1}{j_2}{j_3}{k}{l}{m}\,.
\ee
The latter is equivalent to eq.\ (6), p.454, in \cite{Varshalovich}.

\subsection{Derivation of dual representation}

The transformation to the dual representation proceeds in two steps. First we expand the plaquette action for each face into characters:
\be
\exp\Big(- \clS_f(U_f)\Big) = \sum_{j_f}\;(2j_f+1)\,\e^{-\frac{2}{\beta}\,j_f(j_f + 1)}\,\chi_{j_f}(U_f)
\label{characterexpansion}
\ee
In the second step, we integrate over the connection variable $U_e$, so that there remains only a sum over spin variables.
Before and after the integration, we use, in addition, identity \eq{twostrands} and \eq{extractionidentity}. This allows us to reduce all tensor contractions to $6j$--symbols.

\begin{figure}
\begin{center}
\parbox{5.8cm}{\includegraphics[height=6cm]{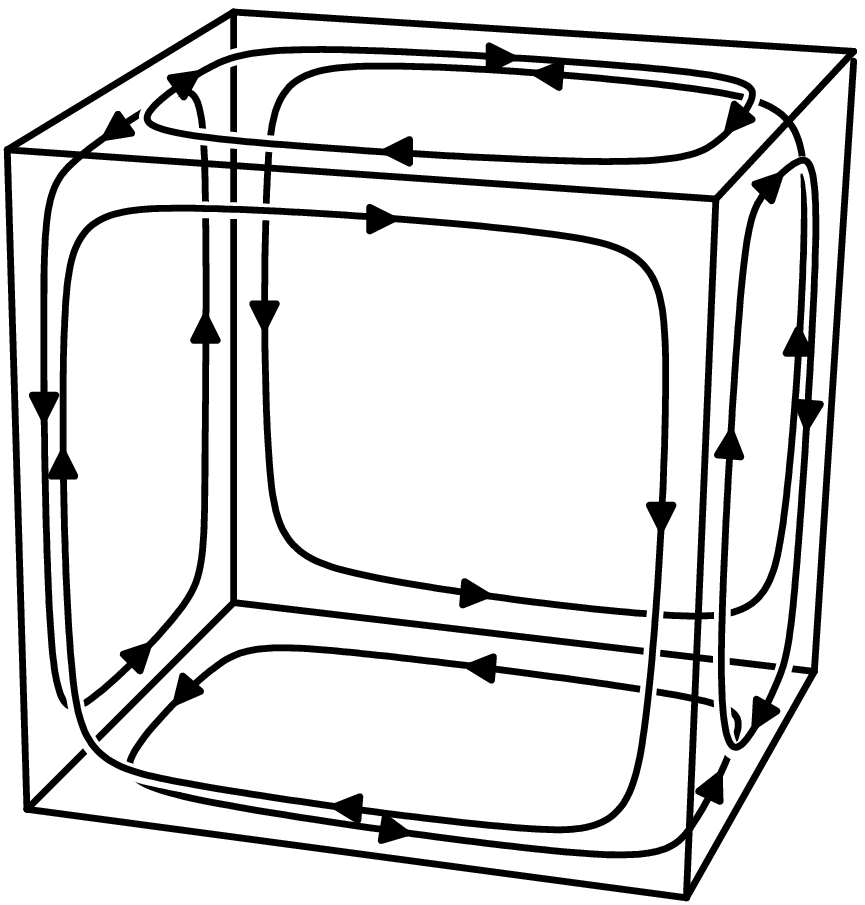}} \qquad $\longrightarrow$ \qquad \parbox{5.8cm}{\includegraphics[height=6cm]{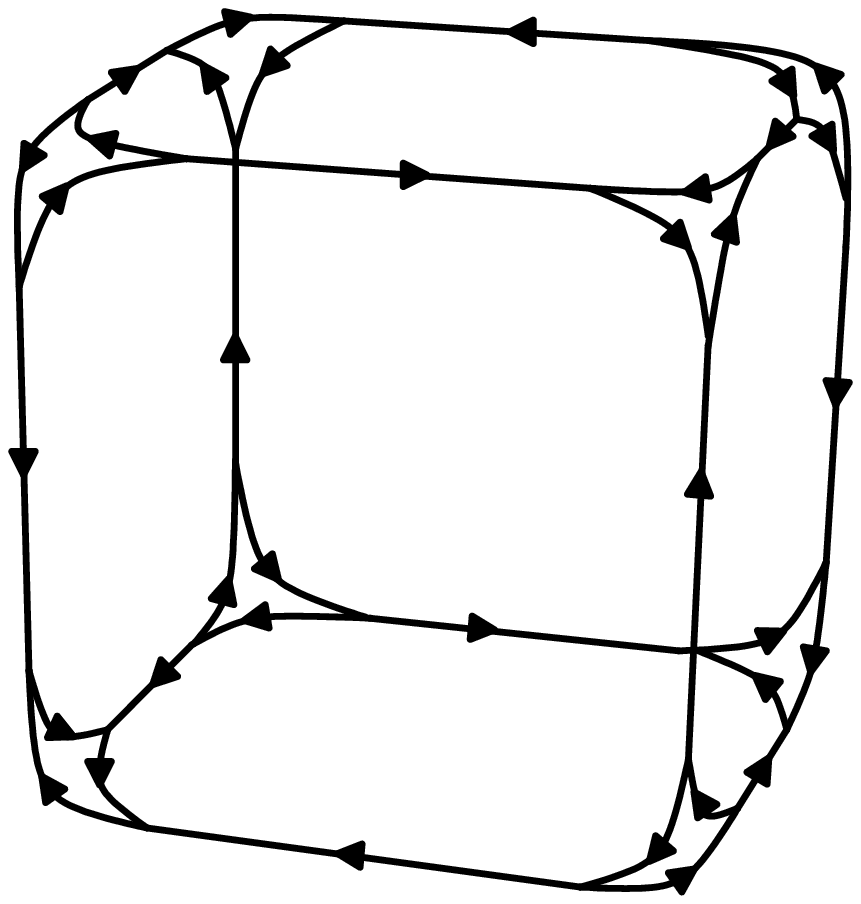}}
\end{center}
\caption{\label{loopsevencube} The loops from the character expansion are attributed to even cubes. 
Application of the identity \eq{twostrands} leads to the diagram on the right--hand side. The loops are drawn without the dots from representation matrices.}
\end{figure}

As explained in the main part, we separate the cubes of the lattice $\kappa$ into alternating even and odd cubes.
Each face of $\kappa$ is attributed to an even cube. In our graphical representation, the trace 
$\chi_{j_f}(U_f)$ in \eq{characterexpansion} becomes a loop of spin $j_f$. Thus, we have a loop of spin $j_f$ for each face $f$ of an even cube (see \fig{loopsevencube}). On each pair of neighbouring strands we can apply identity \eq{twostrands} and thereby turn the six loops into a single diagram.
For each edge of the cube, we get a new strand, a sum over its spin, and the dimension of the representation.

\psfrag{A}{$\sst A$}
\psfrag{B}{$\sst B$}
\psfrag{C}{$\sst C$}
\psfrag{D}{$\sst D$}
\psfrag{E}{$\sst E$}
\psfrag{F}{$\sst F$}
\psfrag{a}{$\sst a$}
\psfrag{b}{$\sst b$}
\psfrag{c}{$\sst c$}
\psfrag{d}{$\sst d$}
\psfrag{e}{$\sst e$}
\psfrag{f}{$\sst f$}
\psfrag{g}{$\sst g$}
\psfrag{h}{$\sst h$}
\psfrag{1}{$\sst 1$}
\psfrag{2}{$\sst 2$}
\psfrag{3}{$\sst 3$}
\psfrag{4}{$\sst 4$}
\psfrag{5}{$\sst 5$}
\psfrag{6}{$\sst 6$}
\psfrag{7}{$\sst 7$}
\psfrag{8}{$\sst 8$}
\psfrag{9}{$\sst 9$}
\psfrag{10}{$\sst 10$}
\psfrag{11}{$\sst 11$}
\psfrag{12}{$\sst 12$}
\pic{labelling}{Labelling of faces, edges and vertices of even cubes in $\kappa^*$.}{7cm}{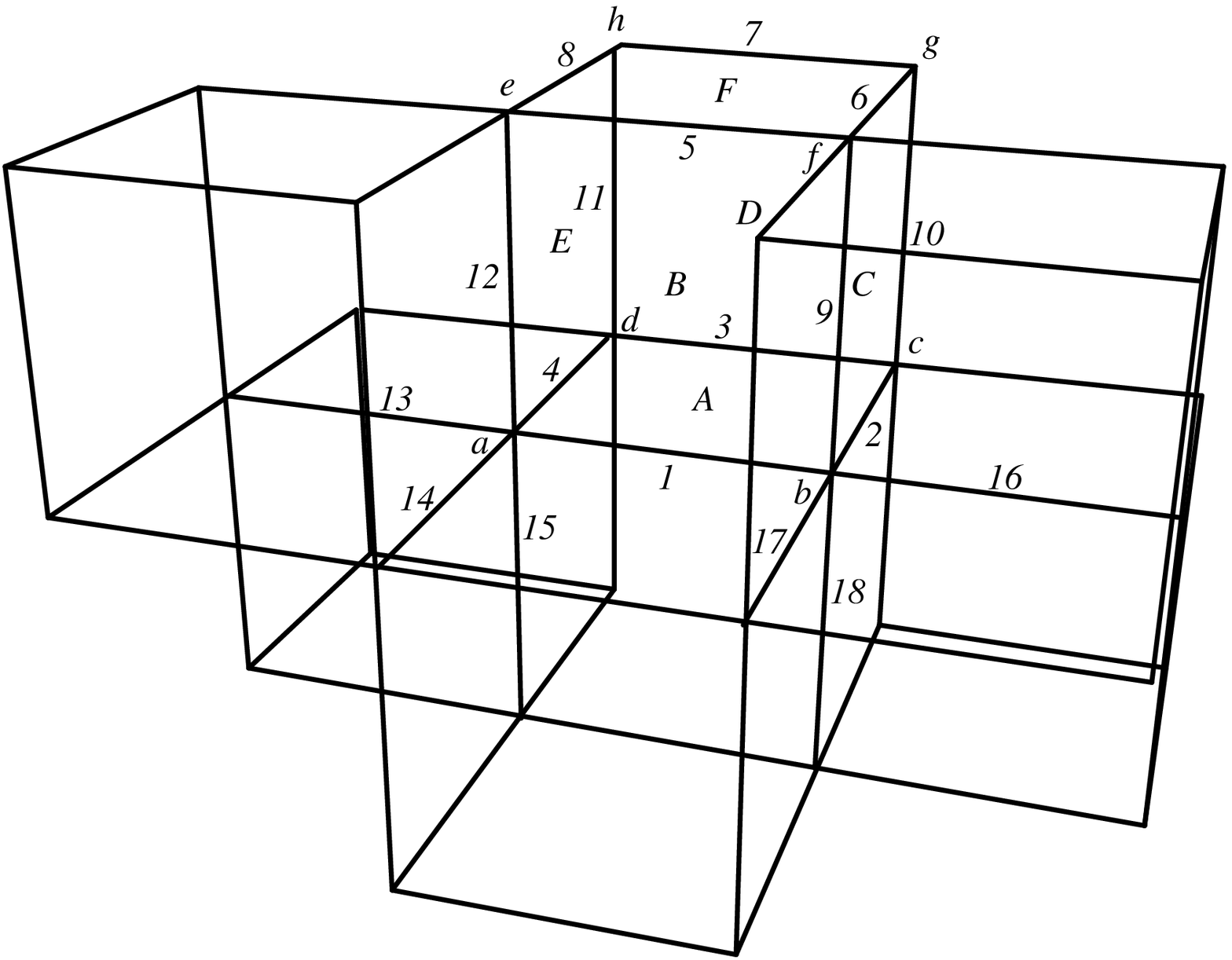}

To proceed further, we need to label the spin variables. We do this by labelling the vertices, edges and faces of the even cube and its immediate vicinity (see \fig{labelling}). As in Diakonov \& Petrov's paper \cite{DiakonovPetrov}, faces are labelled by capital letter $A, B, C \ldots$, edges are given numbers $1, 2, 3 \ldots$, and vertices receive lowercase letters $a, b, c \ldots$ The spins of loops are denoted by $J_X$ where $X$ is the label of the face to which the loop belongs. For each edge $i$, we get two new spins, coming from two even cubes that share the edge: we call them $j_i$ and $j'_i$.
Magnetic quantum numbers are associated to vertices and designated by letters $o, p, q \ldots$ with a subscript for the relevant vertex. 

Thus, we obtain the following, labelled diagram for the even cube:
\be
\left(\prod_i\;(2j_i+1)\right)\times\,\parbox{6.6cm}{\includegraphics[height=6.5cm]{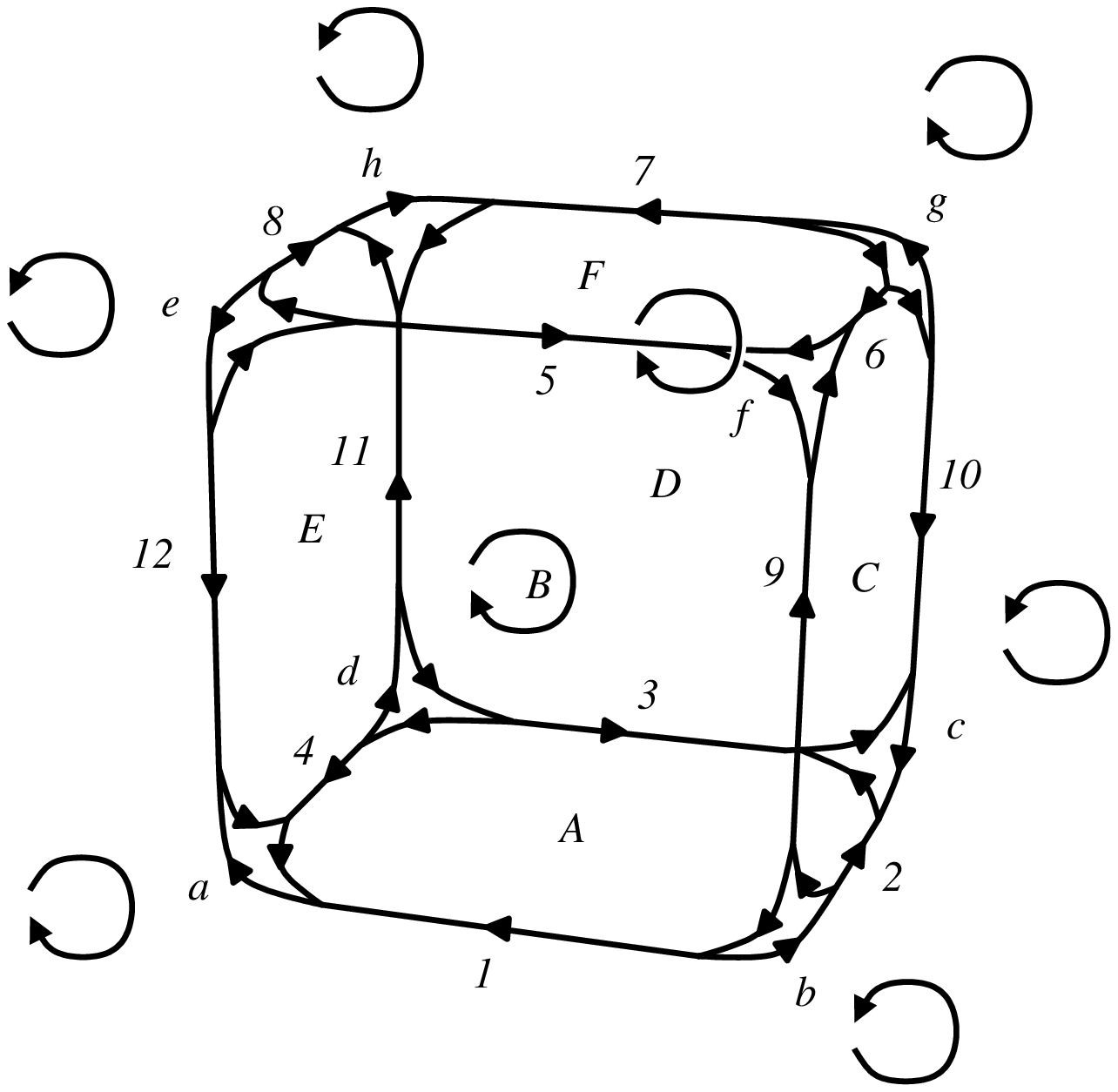}}
\label{diagramafterPeterWeyl}
\ee
Strands along edges carry spins of type $j_i$, while the strands in the corners have the spin $J_X$ of the corresponding face.
Identity \eq{twostrands} requires that nodes on opposite ends have opposite orientation. We satisfy this by choosing a common orientation for all three nodes near a vertex and by letting this orientation alternate as we go from corner to corner. In this manner, we receive two kinds of vertices: vertices of \textit{type I} that have clockwise orientation from outside the cube ($a, c, f, h$), and vertices of \textit{type II} with anti--clockwise orientation from outside the cube ($b, d, g, e$).
We indicate these orientations by eight circles: four of them are situated in front of the cube, and four of them lie behind it. 
This pattern of orientations is consistently extended to all even cubes. Then, a given vertex of $\kappa$ is either of type I for all surrounding even cubes,
or of type II.

Next we apply identity \eq{extractionidentity} to replace the corners of the diagram \eq{diagramafterPeterWeyl} by $6j$--symbols and $3jm$--symbols.
For a type I vertex such as $a$, we get
\psfrag{1}{$1$}
\psfrag{12}{$12$}
\psfrag{4}{$4$}
\psfrag{E}{$E$}
\psfrag{B}{$B$}
\psfrag{A}{$A$}
\bea
\parbox{4.3cm}{\includegraphics[height=4cm]{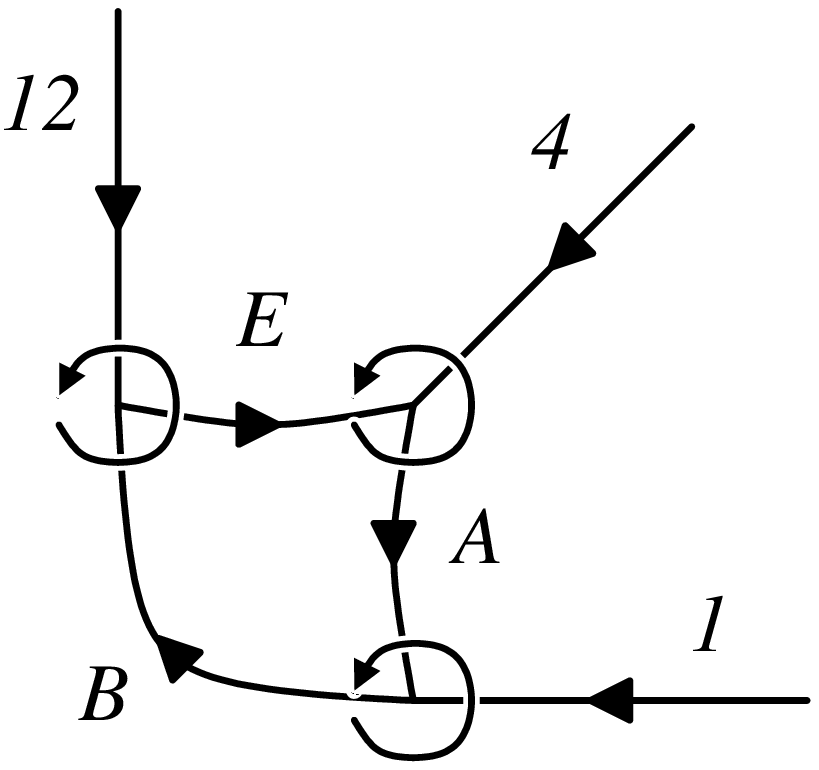}} &\quad=\quad&  \parbox{5.3cm}{\includegraphics[height=3.2cm]{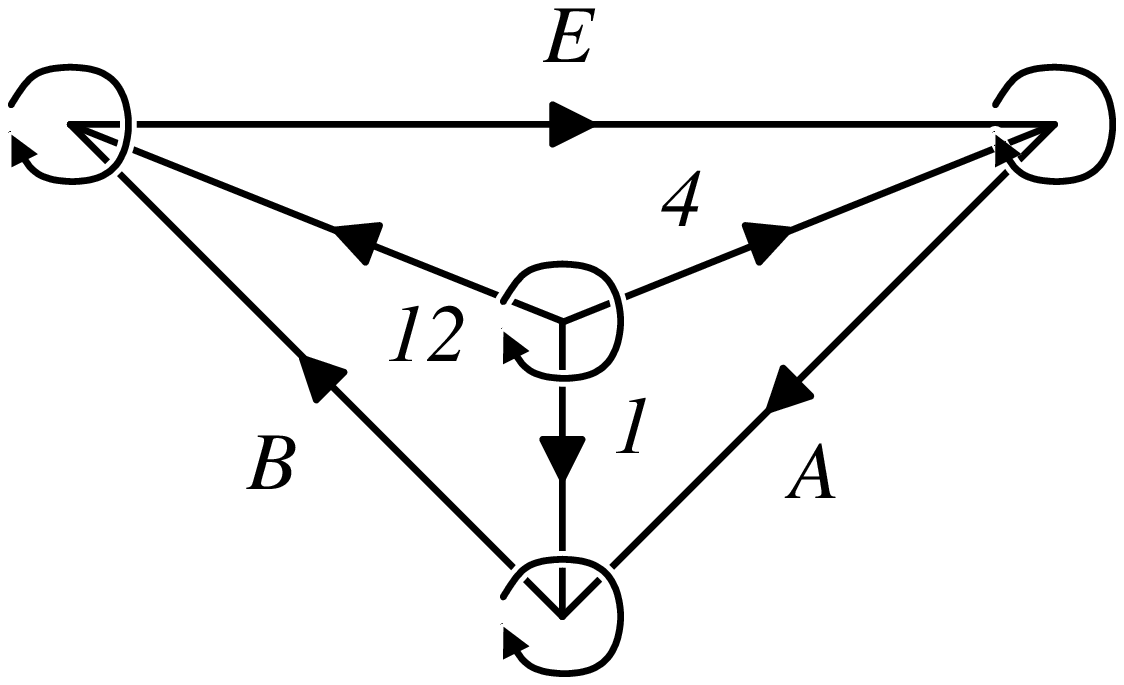}}
\quad\parbox{3.3cm}{\includegraphics[height=2.2cm]{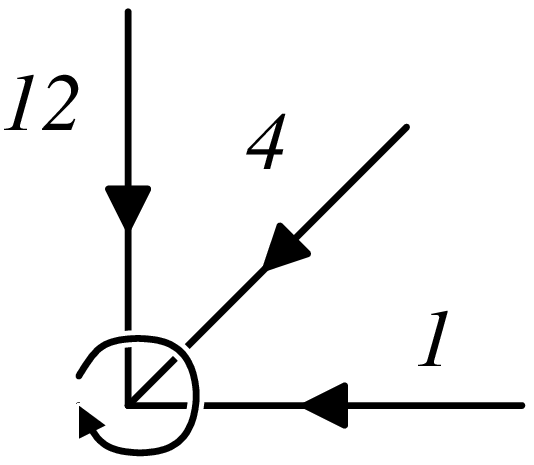}} \\
&\quad=\quad& \sixj{j_1}{j_{12}}{j_4}{J_E}{J_A}{J_B}\quad\parbox{3.3cm}{\includegraphics[height=2.2cm]{33_3jmextractionvertexI}}\,.
\eea
For a type II vertex like $b$, we have, on the other hand,
\psfrag{1}{$1$}
\psfrag{2}{$2$}
\psfrag{9}{$9$}
\psfrag{C}{$C$}
\bea
\lefteqn{\parbox{5.7cm}{\includegraphics[height=3.5cm]{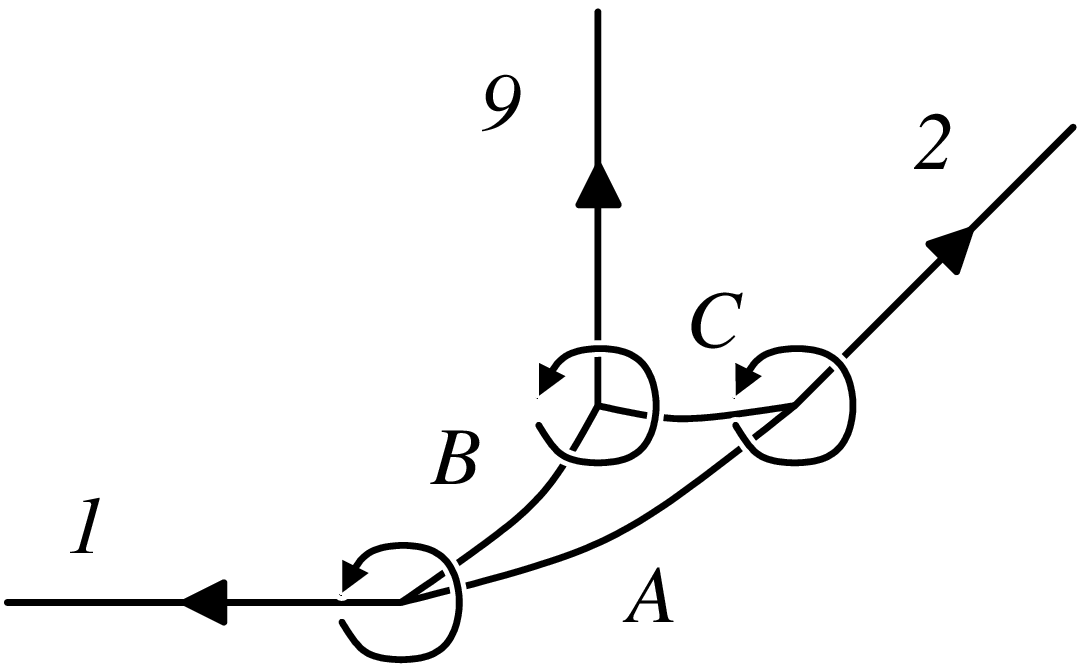}} =  \parbox{5.3cm}{\includegraphics[height=3.2cm]{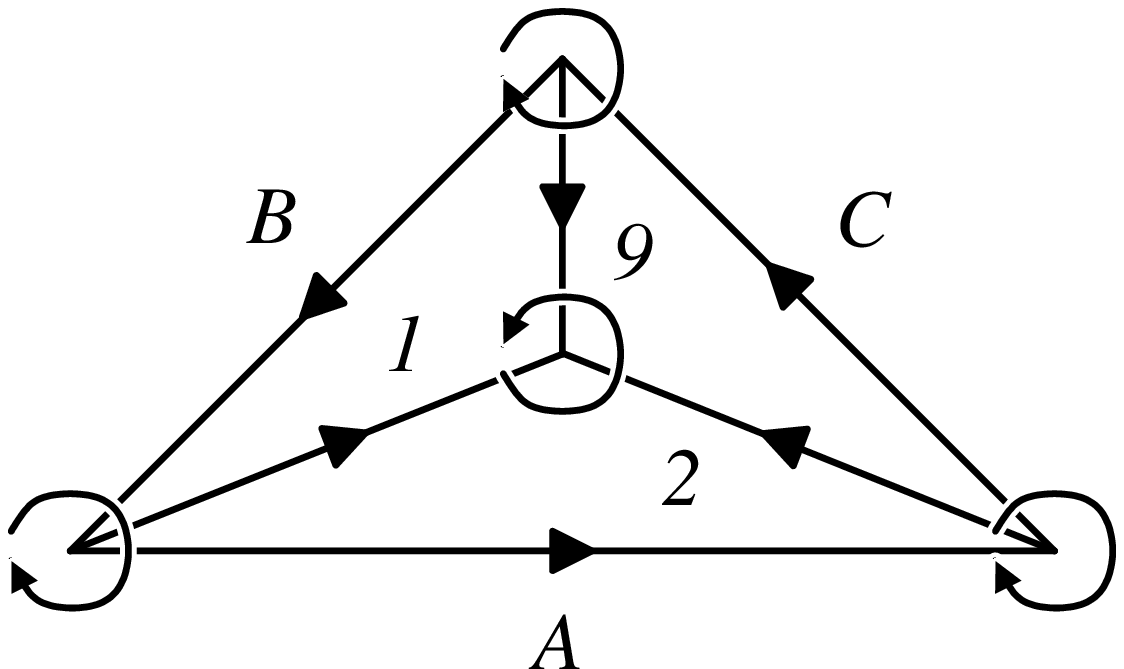}}
\quad\parbox{3.3cm}{\includegraphics[height=2.2cm]{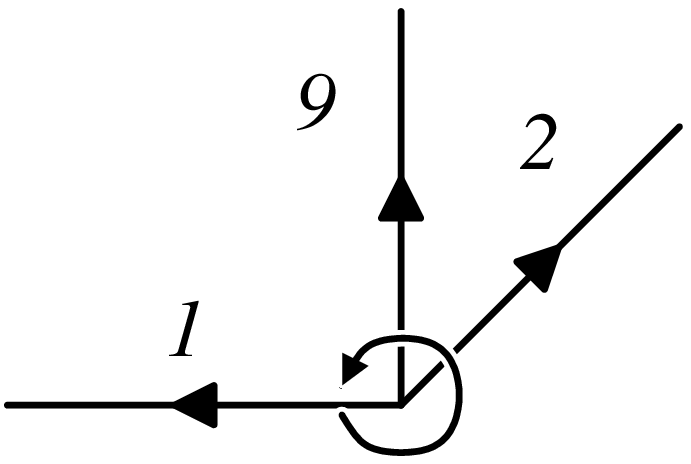}}} \nonumber \\
&\quad=\quad&  (-1)^{j_1 + j_2 + j_9}\parbox{5.3cm}{\includegraphics[height=3.2cm]{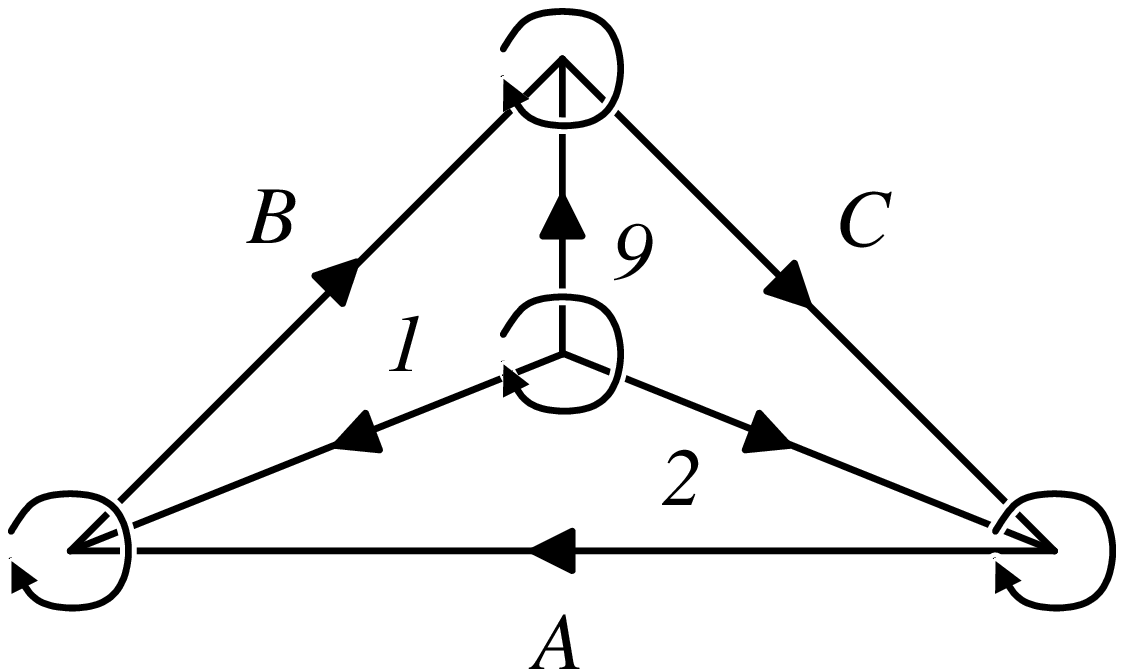}}
\quad\parbox{3.3cm}{\includegraphics[height=2.2cm]{36_3jmextractionvertexII}} \\
&\quad=\quad& (-1)^{j_1 + j_2 + j_9}\sixj{j_1}{j_2}{j_9}{J_C}{J_B}{J_A}\quad\parbox{3.3cm}{\includegraphics[height=2.2cm]{36_3jmextractionvertexII}}
\eea
We repeat this procedure everywhere in the lattice and obtain a $6j$-- and $3jm$-- for every corner of an even cube.
Altogether this gives us four $6j$-- and $3jm$--symbols for each vertex. For each edge, we receive, in addition, a sign factor $(-1)^{j_i + j'_i}$ and a factor $(2j_i + 1)(2j'_i+1)$.

\psfrag{I}{$\mathrm{I}$}
\psfrag{II}{$\mathrm{II}$}
\psfrag{a}{$a$}
\psfrag{b}{$b$}
\psfrag{c}{$c$}
\psfrag{d}{$d$}
\psfrag{e}{$e$}
\psfrag{f}{$f$}
\psfrag{g}{$g$}
\psfrag{h}{$h$}
\psfrag{1}{$1$}
\psfrag{1'}{$1'$}
\psfrag{2}{$2$}
\psfrag{2'}{$2'$}
\psfrag{3}{$3$}
\psfrag{4}{$4$}
\psfrag{5}{$5$}
\psfrag{6}{$6$}
\psfrag{7}{$7$}
\psfrag{8}{$8$}
\psfrag{9}{$9$}
\psfrag{10}{$10$}
\psfrag{11}{$11$}
\psfrag{12}{$12$}
\psfrag{13}{$13$}
\psfrag{14}{$14$}
\psfrag{14'}{$14'$}
\psfrag{15}{$15$}
\psfrag{16}{$16$}
\psfrag{17}{$17$}
\psfrag{18}{$18$}
\pic{6jjextraction}{Contraction of $3jm$--symbols: the dots indicate representation matrices $D_j(U_i)$. If the Polaykov loop passes through the vertices $a$ and $b$, there is an additional strand of spin $j$ (drawn in the middle between $3jm$--symbols).}{7cm}{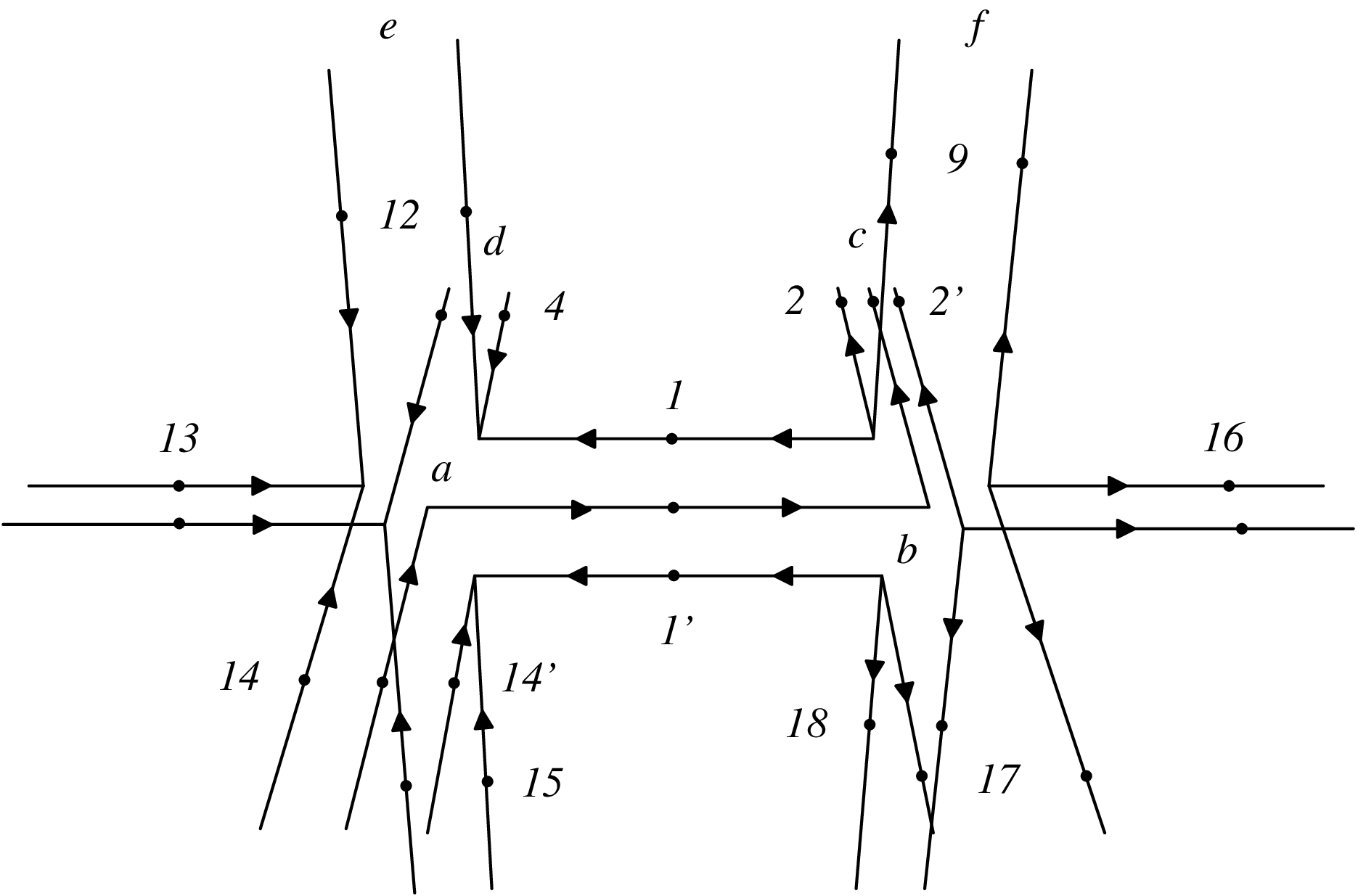}
\psfrag{1'}{$1'$}
\psfrag{2'}{$2'$}
\psfrag{14'}{$14'$}
\psfrag{a}{$a$}
\psfrag{b}{$b$}

The $3jm$--symbols at a vertex are connected to those at other vertices by representation matrices (see \fig{6jjextraction}). We integrate over these representation matrices using identity \eq{integral2strands} and \eq{integral3strands}. The result is a contraction of tensors for each vertex. 
When $a$ and $b$ are two vertices in the Polyakov loop $C$, we obtain
\[
\parbox{13.7cm}{\includegraphics[height=4.5cm]{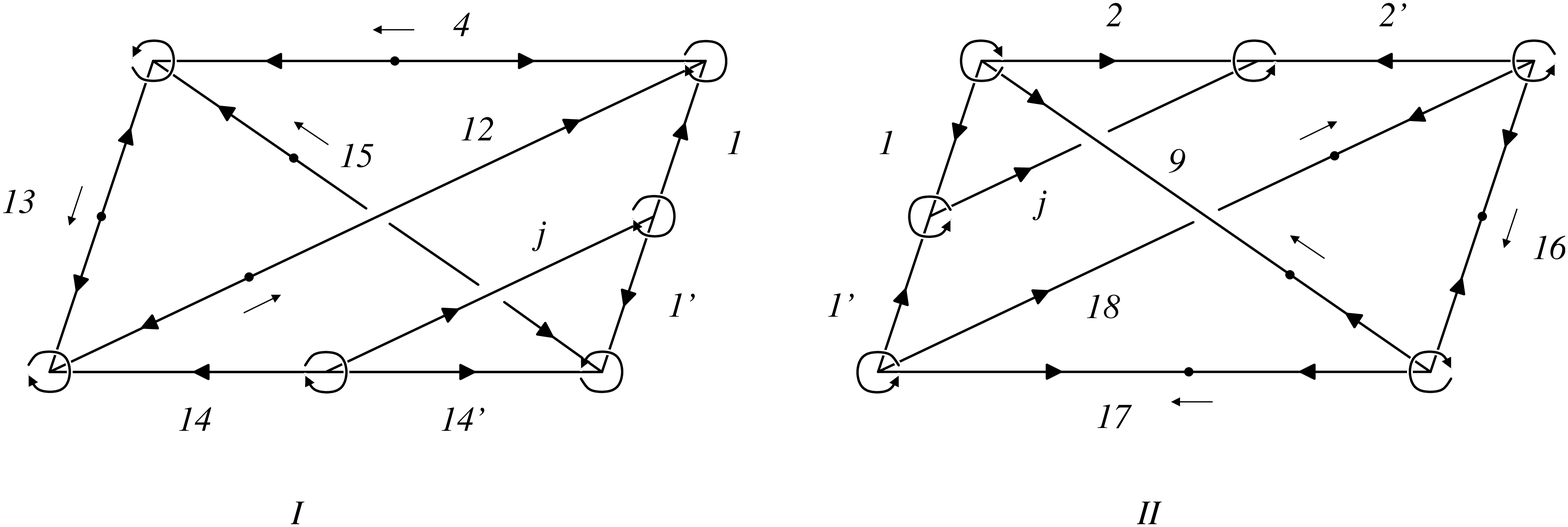}}\\,,
\label{diagramtypeIandII}
\]
For every edge outside $C$, the integration forces $j_i$ and $j'_i$ to be the same and gives, in addition, a factor $(2j_i + 1)^{-1}$. 

It is sufficient to evaluate these two spin networks, since the path of the Polyakov loop is periodic. The spin networks outside the Polyakov loop are the same with the edges of spin $j$ deleted. As we see below, the spin network at $a$ (type I) is related to that at $b$ (type II) by a number of simple operations. Thus, we only need to compute the spin network of type II. We use identity \eq{extractionidentity} to split it into two diagrams with 6 spins\footnote{This simple splitting of diagrams is possible, since we have chosen a zig--zag path for the Polyakov loop. For a straight Wilson line, we would receive a $9j$--symbol. In that case, the reduction to $6j$--symbols is more complicated, and the asymptotic behaviour not known.}:
\psfrag{II1}{$\mathrm{II}_1$}
\psfrag{II2}{$\mathrm{II}_2$}
\psfrag{o}{$\sst o$}
\psfrag{p}{$\sst p$}
\psfrag{v}{$\sst v$}
\psfrag{u}{$\sst u$}
\psfrag{w}{$\sst w$}
\psfrag{m}{$\sst m$}
\psfrag{s}{$\sst s$}
\psfrag{q}{$\sst q$}
\psfrag{r}{$\sst r$}
\psfrag{t}{$\sst t$}
\psfrag{x}{$\sst x$}
\psfrag{y}{$\sst y$}
\psfrag{z}{$\sst z$}
\be
\parbox{14.3cm}{\includegraphics[height=4.8cm]{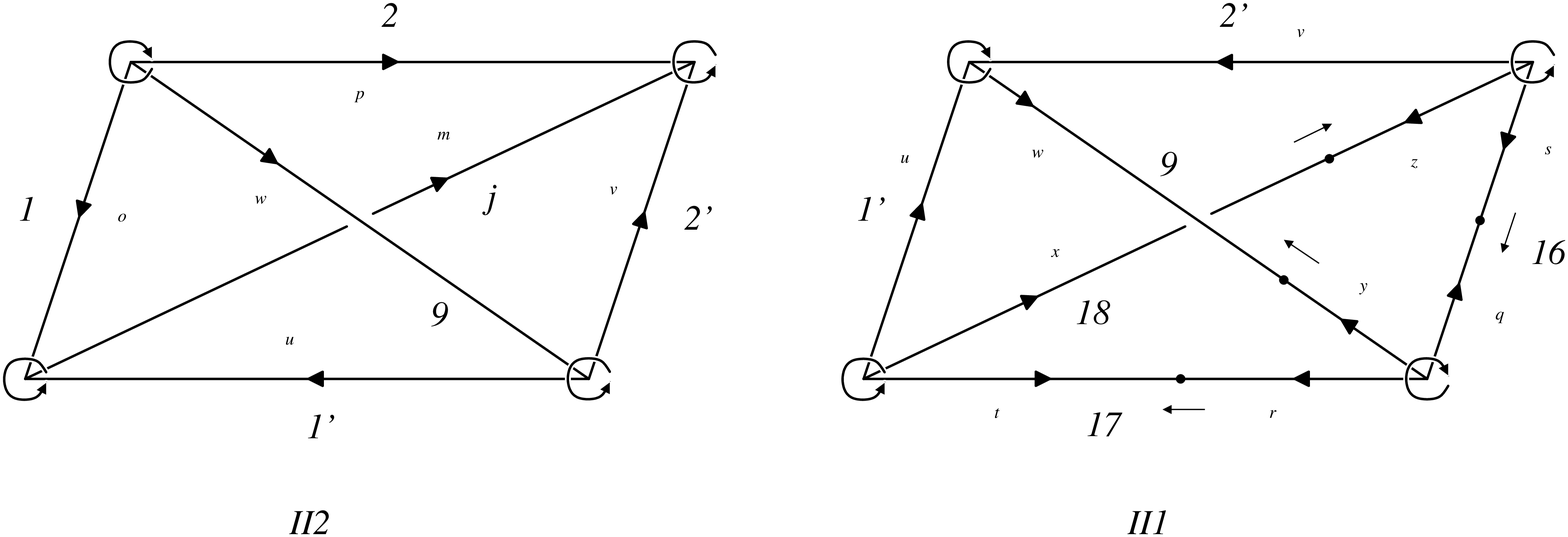}}
\label{diagramII1andII2}
\ee
The modulus of these diagrams is equal to $6j$--symbols. It remains to compute their phase.
We do this by translating the graphs back into formulas. For that purpose, we have labelled each edge by lowercase letters to indicate the magnetic quantum number of the $3jm$--symbols. For diagram $\mathrm{II}_1$, we receive
\bea
\mathrm{II}_1
&=& 
(-1)^{j_9 + y_b}\,\delta_{y_b,-w_b}(-1)^{j_9 + w_b}\threejm{j'_1}{j'_2}{j_9}{u_b}{v_b}{-w_b}\,
(-1)^{j_{17} + r_b}\threejm{j'_1}{j_{17}}{j_{18}}{u_b}{-r_b}{x_b} \nonumber \\
&& \times\,
(-1)^{j_{18} + x_b}\threejm{j'_2}{j_{18}}{j_{16}}{v_b}{-x_b}{s_b}\,
(-1)^{j_{16} + s_b}\threejm{j_9}{j_{16}}{j_{17}}{y_b}{-s_b}{r_b} \nonumber \\
&=&
(-1)^{2j_9}\threejm{j'_1}{j'_2}{j_9}{-u_b}{-v_b}{-y_b}\,
(-1)^{j_{18} + x_b}\threejm{j'_1}{j_{17}}{j_{18}}{-u_b}{-r_b}{x_b} \nonumber \\
&& \times\,
(-1)^{j_{16} + s_b}\threejm{j'_2}{j_{18}}{j_{16}}{-v_b}{-x_b}{s_b}\,
(-1)^{j_{17} + r_b}\threejm{j_9}{j_{16}}{j_{17}}{-y_b}{-s_b}{r_b} \nonumber \\
&=&
(-1)^{2j_9 + \sum_i 2j_i}\threejm{j'_1}{j'_2}{j_9}{u_b}{v_b}{y_b}\,
(-1)^{j_{18} + x_b}\threejm{j'_1}{j_{17}}{j_{18}}{u_b}{r_b}{-x_b} \nonumber \\
&& \times\,
(-1)^{j_{16} + s_b}\threejm{j'_2}{j_{18}}{j_{16}}{v_b}{x_b}{-s_b}\,
(-1)^{j_{17} + r_b}\threejm{j_9}{j_{16}}{j_{17}}{y_b}{s_b}{-r_b} \nonumber \\
&=&
(-1)^{2j_9 + \sum_i 2j_i}\sixj{j'_1}{j'_2}{j_9}{j_{16}}{j_{17}}{j_{18}} = (-1)^{2j_9 + 2j'_2 + 2j_{17}}\sixj{j'_1}{j'_2}{j_9}{j_{16}}{j_{17}}{j_{18}} \nonumber \\
&=& (-1)^{2j'_1 + 2j_{17}}\sixj{j'_1}{j'_2}{j_9}{j_{16}}{j_{17}}{j_{18}}\,.
\eea
Diagram $\mathrm{II}_2$ gives
\bea
\mathrm{II}_2
&=& 
(-1)^{j'_1 + j_9 + j'_2}\threejm{j_1}{j_2}{j_9}{o_b}{p_b}{w_b}\,
(-1)^{j + m_b}\threejm{j_1}{j'_1}{j}{o_b}{u_b}{-m_b} \nonumber \\
&& \times\,
\threejm{j_2}{j}{j'_2}{p_b}{m_b}{v_b}\,
(-1)^{j_9 + w_b}\threejm{j_9}{j'_2}{j'_1}{-w_b}{v_b}{u_b} \nonumber \\
&=&
(-1)^{j'_1 + j_9 + j'_2}
\threejm{j_1}{j_2}{j_9}{o_b}{p_b}{w_b}\,
(-1)^{j + m_b}\threejm{j_1}{j'_1}{j}{o_b}{u_b}{-m_b} \nonumber \\
&& \times\,
\threejm{j_2}{j}{j'_2}{p_b}{m_b}{-v_b}\,
(-1)^{j_9 + w_b}\threejm{j_9}{j'_2}{j'_1}{-w_b}{-v_b}{u_b} \nonumber \\
&=&
(-1)^{j'_2 + v_b + j_9 + w_b + j'_1 + u_b}
\threejm{j_1}{j_2}{j_9}{o_b}{p_b}{w_b}\,
(-1)^{j + m_b}\threejm{j_1}{j'_1}{j}{o_b}{u_b}{-m_b} \nonumber \\
&& \times\,
(-1)^{j'_2 + v_b}\threejm{j_2}{j}{j'_2}{p_b}{m_b}{-v_b}\,
(-1)^{j'_1 + u_b}\threejm{j_9}{j'_2}{j'_1}{-w_b}{-v_b}{u_b} \nonumber \\
&=&
(-1)^{j'_2 + j_9 + j'_1 + 2u_b}\sixj{j_1}{j_2}{j_9}{j'_2}{j'_1}{j} = (-1)^{j'_1 + j'_2 + j_9 + 2j'_1}\sixj{j'_1}{j'_2}{j_9}{j_2}{j_1}{j}\,.
\eea
Together this yields
\bea
\mathrm{II} = \mathrm{II}_2\mathrm{II}_1 &=& 
(-1)^{j'_1 + j'_2 + j_9 + 2j'_1}\sixj{j'_1}{j'_2}{j_9}{j_2}{j_1}{j}\,(-1)^{2j'_1 + 2j_{17}}\sixj{j'_1}{j'_2}{j_9}{j_{16}}{j_{17}}{j_{18}} \\
&=& 
(-1)^{j'_1 + j'_2 + j_9 + 2j_{17}}\sixj{j'_1}{j'_2}{j_9}{j_2}{j_1}{j}\sixj{j'_1}{j'_2}{j_9}{j_{16}}{j_{17}}{j_{18}}\,.
\label{spinnetworkII}
\eea
The spin network of vertex $a$ (type I) is related to that of vertex $b$ (type II) by the following steps:
\begin{enumerate}
\item rotate the lower right corner of spin network II onto the upper left corner and identify its edges with those of spin network I,
\item take the complex conjugate (i.e.\ reverse all arrows on edges),
\item reverse the arrow on the node of edge 15,
\item reverse the arrow on the edge $j$. 
\end{enumerate}
Step 1.\ corresponds to a relabelling of spins in \eq{spinnetworkII}. Step 2.\ does not change the value, since we deal with real quantities. In step 3.\ we get a factor $(-1)^{2j_{15}}$, since
\be
\threejm{j_{15}}{0}{j_{15}}{w_a}{0}{y_a} = (-1)^{2j_{15}}\threejm{j_{15}}{0}{j_{15}}{y_a}{0}{w_a}
\ee
while step 4.\ does not produce any factors:
\bea
&& \sum_{m_a}\; (-1)^{j+m_a} \threejm{j_{14}}{j}{j'_{14}}{p_a}{-m_a}{v_a}\, (-1)^{j+m_a} \threejm{j'_1}{j}{j_1}{u_a}{-m_a}{o_a} \nonumber \\
&&= \quad
\sum_{m_a}\; \threejm{j_{14}}{j}{j'_{14}}{p_a}{m_a}{v_a}\threejm{j'_1}{j}{j_1}{u_a}{m_a}{o_a}
\eea
Therefore,
\bea
\mathrm{I} &=& (-1)^{2j_{15}}\,\mathrm{II}_{\mathrm{relabelled}} \\
&=& (-1)^{2j_{15}}\,(-1)^{j_{14} + j_1 + j_{15} + 2j_{13}}\sixj{j_{14}}{j_1}{j_{15}}{j'_1}{j'_{14}}{j}\sixj{j_{14}}{j_1}{j_{15}}{j_4}{j_{13}}{j_{12}} \\
&=& (-1)^{j_1 + j_{14} + j_{15} + 2j_4}\sixj{j_1}{j_{14}}{j_{15}}{j'_{14}}{j'_1}{j}\sixj{j_1}{j_{14}}{j_{15}}{j_{13}}{j_4}{j_{12}}
\eea
In summary: along the Polyakov loop, vertices of type I and II contribute
\bea 
\mathrm{I} &=& (-1)^{j_1 + j_{14} + j_{15} + 2j_4}\sixj{j_1}{j_{14}}{j_{15}}{j'_{14}}{j'_1}{j}\sixj{j_1}{j_{14}}{j_{15}}{j_{13}}{j_4}{j_{12}}\,, \\
\mathrm{II} &=& (-1)^{j'_1 + j'_2 + j_9 + 2j_{17}}\sixj{j'_1}{j'_2}{j_9}{j_2}{j_1}{j}\sixj{j'_1}{j'_2}{j_9}{j_{16}}{j_{17}}{j_{18}}\,.
\eea
For vertices outside the Polyakov loop, this reduces to
\bea 
\mathrm{I|_{j=0}} &=& \frac{1}{\sqrt{2j_{14} + 1}\sqrt{2j_1 + 1}}\, (-1)^{2j_4}\sixj{j_1}{j_{14}}{j_{15}}{j_{13}}{j_4}{j_{12}}\,, \\
\mathrm{II|_{j=0}} &=& \frac{1}{\sqrt{2j_1 + 1}\sqrt{2j_2 + 1}}\, (-1)^{2j_{17}}\sixj{j_1}{j_2}{j_9}{j_{16}}{j_{17}}{j_{18}}\,.
\eea
Since vertices of type I and II alternate along the lattice, the factors of $(-1)^{2j_4}$ and $(-1)^{2j_{17}}$ cancel each other, when we multiply all vertex factors. Outside the Polyakov loop, the $\sqrt{2j_i+1}$ factors combine to give $(2j_i+1)^{-1}$. 

Let us summarize what we have obtained so far: the transition of \fig{loopsevencube} introduced new spins $j_i$ and $j'_i$ and factors $(2j_i+1)(2j'_i+1)$ for each edge. From the application of identity \eq{extractionidentity} to diagram \eq{diagramafterPeterWeyl}, we got a factor
\be
\sixj{j_k}{j_l}{j_m}{J_X}{J_Y}{J_Z}
\ee
for each even cube near a vertex. Due to type II vertices, we also had a sign factors $(-1)^{j_i + j'_i}$ per edge. The integration over group variables
produced factors
\bea 
\mathrm{I} &=& (-1)^{j_1 + j_{14} + j_{15}}\sixj{j_1}{j_{14}}{j_{15}}{j'_{14}}{j'_1}{j}\sixj{j_1}{j_{14}}{j_{15}}{j_{13}}{j_4}{j_{12}}\,, \\
\mathrm{II} &=& (-1)^{j'_1 + j'_2 + j_9}\sixj{j'_1}{j'_2}{j_9}{j_2}{j_1}{j}\sixj{j'_1}{j'_2}{j_9}{j_{16}}{j_{17}}{j_{18}}\,.
\eea
for vertices along the Polyakov loop $C$, and outside of it
\bea 
\mathrm{I} &=& \sixj{j_1}{j_{14}}{j_{15}}{j_{13}}{j_4}{j_{12}}\,, \\
\mathrm{II} &=& \sixj{j_1}{j_2}{j_9}{j_{16}}{j_{17}}{j_{18}}\,.
\eea
For every edge outside $C$, the integration imposes $j_i = j'_i$ and gives a factor $(2j_i + 1)^{-1}$. 

We now switch to the description in terms of the modified triangulation $\Tt$: spins $J_X$ on faces $f\subset\kappa$ turn into spins $j_e$ on edges $e\subset\kappa^*$, and spins $j_i$ and $j'_i$ on edges become spins $j_e$ on diagonals in $\Tt$. The spin foam sum takes the form
\bea
\b \tr_j U_C\ket &=& \frac{1}{Z}\,\sum_{\{j_e\}_{\Tt}}
\left(\prod_{e\subset \Tt} (2j_e+1)\right) \nonumber \\
&& \times\,
\left(\prod_{\mbox{\tiny single diag.\ $e \subset \Tt$}} (-1)^{2j_e}\right) 
\left(\prod_{\mbox{\tiny double diag.\ $e, e' \subset \Tt$}} (-1)^{j_e + j_{e'}}\right) 
\left(\prod_{t\subset \Tt} A_t\right)
\left(\prod_{e\subset\kappa^*}\,\e^{-\frac{2}{\beta}\,j_e(j_e + 1)}\right)\,,
\eea
where for tetrahedra as in \fig{TandTprime}a
\be
A_t = \sixj{j_1}{j_2}{j_3}{j_4}{j_5}{j_6}\,,
\ee
and for degenerate tetrahedra (as in \fig{TandTprime}b)
\be
A_t = (-1)^{j_1 + j_2 + j_3 + j}\,\sixj{j_1}{j_3}{j_2}{j'_3}{j'_1}{j}\,.
\ee
Except for sign factors these formulas are identical to eq.\ \eq{spinfoamsumPolyakovloop}, \eq{amplitudetetrahedron} and \eq{amplitudedegeneratetetrahedron}.

To show equivalence with \eq{spinfoamsumPolyakovloop}, we need to shift the sign factors on diagonals to edges of $\kappa^*$.
For this purpose, we think of the triangulation as being made up of octahedra around vertices of $\kappa^*$:
\psfrag{A}{$\ssst A$}
\psfrag{B}{$\ssst B$}
\psfrag{C}{$\ssst C$}
\psfrag{D}{$\ssst D$}
\psfrag{E}{$\ssst E$}
\psfrag{F}{$\ssst F$}
\psfrag{G}{$\ssst G$}
\psfrag{H}{$\ssst H$}
\psfrag{1}{$\ssst 1$}
\psfrag{2}{$\ssst 2$}
\psfrag{3}{$\ssst 3$}
\psfrag{4}{$\ssst 4$}
\psfrag{5}{$\ssst 5$}
\psfrag{6}{$\ssst 6$}
\psfrag{7}{$\ssst 7$}
\psfrag{8}{$\ssst 8$}
\psfrag{9}{$\ssst 9$}
\psfrag{10}{$\ssst 10$}
\psfrag{11}{$\ssst 11$}
\psfrag{12}{$\ssst 12$}
\[
\parbox{5cm}{\includegraphics[height=5cm]{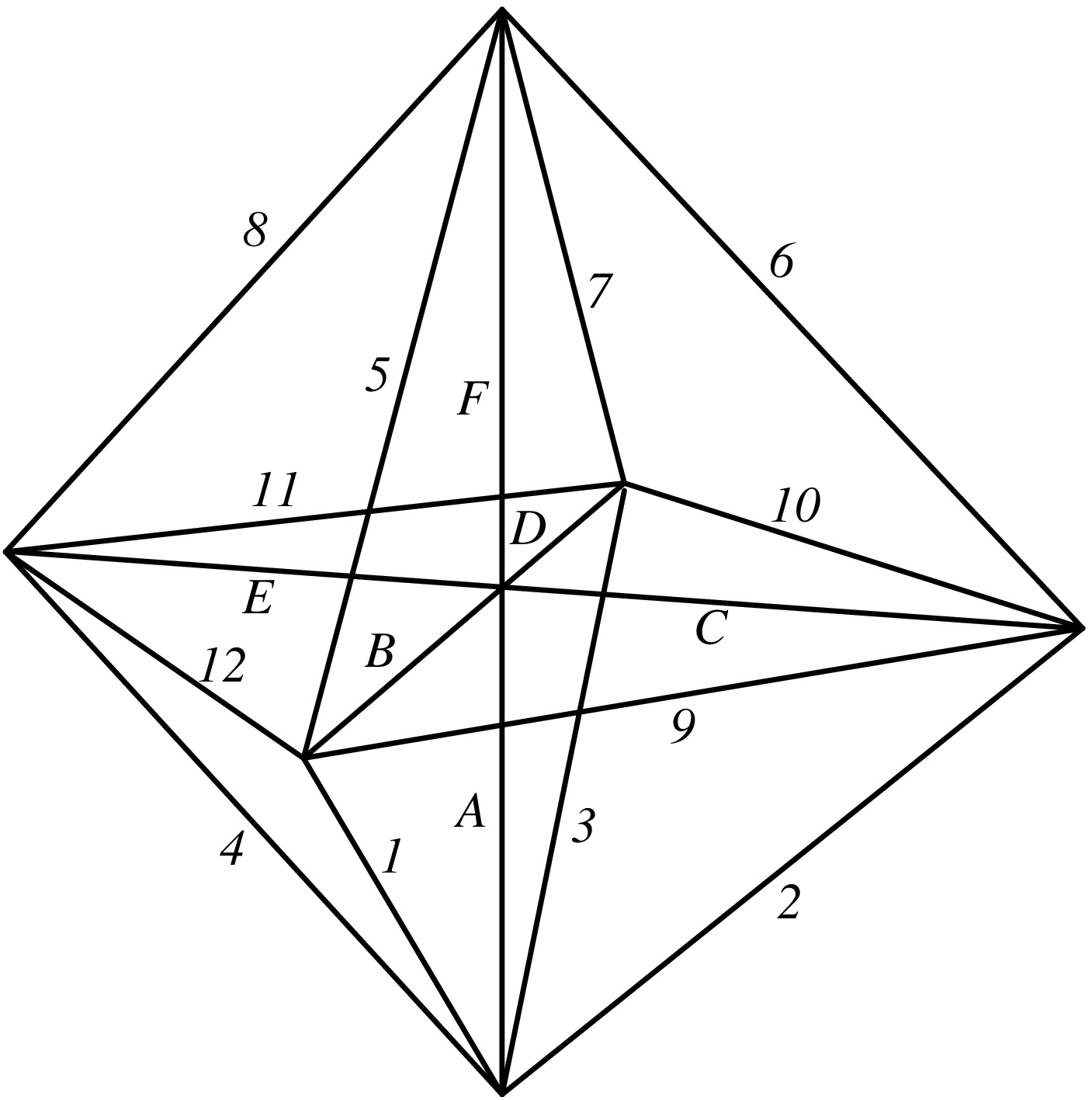}}
\]
Let us first regard the case, where the Polyakov loop is absent. Then, $\Tt$ is the same as $T$ and we have a sign factor $(-1)^{2j_e}$ for each diagonal $e$ of $T$. 
We can distribute these factors such that each factor belongs to exactly one octehadron. One possibility is, for example, that every octehadron
carries the factors
\bea
\lefteqn{(-1)^{2j_9 + 2j_6 + 2j_8 + 2j_{12} + 2j_7 + 2j_3} = (-1)^{2J_A + 2J_F}} \\
&=&  (-1)^{2J_A + 2J_B + 2J_C + 2J_D + 2J_E + 2J_F}\,(-1)^{2j_{12} + 2j_{10}}   
\eea
When we multiply the octehadra, the factors $(-1)^{2j_{12} + 2j_{10}}$ cancel each other due to periodicity.
Therefore, we have the identity
\be
(-1)^{\sum_{e\nsubseteq\kappa^*} 2j_e} = (-1)^{\sum_{e\subset\kappa^*} 2j_e}\,,
\ee
i.e.\ the sign factors can be shifted from diagonals of $T$ to edges of $\kappa^*$.

In the presence of the Polyakov loop, the argument is slightly modified. Depending on its location, the Polyakov loop affects edges of type 8 and 7 or 1 and 2. Suppose the Polyakov loop goes through $8$ and $7$.
Then, we have factors
\be
(-1)^{j_8 + j'_8}\,(-1)^{j_7 + j'_7} = (-1)^{2j'_8 + 2j'_7}\times (-1)^{j_8 - j'_8}\,(-1)^{j_7 - j'_7} 
\ee
from the double edges 8 and 7. We repeat our previous argument with the first factor on the right-hand side, and absorb the second factor into the tetrahedral 
amplitude. When the Polyakov loop passes through 1 and 2, nothing changes, since the factors of 1 and 2 are attributed to adjacent octahedrons.
This gives us precisely the amplitude we described in the main part of the paper.

\end{appendix}

\bibliography{bibliography}
\bibliographystyle{hunsrt}  

\end{document}